\documentclass[letterpaper]{article}
\usepackage[preprint]{aaai2027}
\usepackage[hyphens]{url}
\usepackage{graphicx}
\usepackage{natbib}
\usepackage{caption}
\usepackage{algorithm}
\usepackage{algorithmic}
\usepackage{newfloat}
\usepackage{listings}
\DeclareCaptionStyle{ruled}{labelfont=normalfont,labelsep=colon,strut=off}
\floatstyle{ruled}
\newfloat{listing}{tb}{lst}{}
\floatname{listing}{Listing}
\usepackage{booktabs}
\usepackage{placeins}
\title{DiagChain: A Diagnostic Benchmark for Evaluating LLM Agents on Evidence-Grounded Attack Chain Reconstruction}
\author{
Xuyang Liu\textsuperscript{\rm 1}, Yibin Han\textsuperscript{\rm 2}, Zhenwei Zhang\textsuperscript{\rm 2}, Kai Chang\textsuperscript{\rm 2},\\
Zhiwei Xu\textsuperscript{\rm 1}, Tian Qiu\textsuperscript{\rm 1}, Weixian Deng\textsuperscript{\rm 1}, Jiabao Gao\textsuperscript{\rm 1},\\
Xiaolin Peng\textsuperscript{\rm 1}, Hai Wan\textsuperscript{\rm 1}, Xibin Zhao\textsuperscript{\rm 1}
}
\affiliations{
\textsuperscript{\rm 1}KLISS, BNRist, School of Software, Tsinghua University, Beijing, China\\
\textsuperscript{\rm 2}CRRC Corporation Limited, Beijing, China\\
\textsuperscript{\rm 1}\{liuxuyan23,xzw24,qt24,deng-wx23,gjb25,pxl25\}@mails.tsinghua.edu.cn\\
\textsuperscript{\rm 1}\{wanhai,zxb\}@tsinghua.edu.cn
}

\begin{document}

\maketitle

\begin{abstract}
 
Large Language Model (LLM) agents offer a promising approach to attack chain reconstruction by retrieving and interpreting heterogeneous telemetry to infer ordered attacker actions. However, existing benchmarks mainly evaluate final outputs or aggregate accuracy, providing limited insight into how errors arise and propagate across intermediate reasoning stages. We present \textbf{DiagChain}, a diagnostic benchmark for evidence-grounded attack chain reconstruction that enables stage-wise evaluation of LLM agents. DiagChain includes MAIN-69, a suite of 69 scenarios spanning multiple operating systems, evidence noise levels, and chain lengths. It further introduces Evidence-Centric Retrieval-Augmented Generation (ECRAG), which couples evidence retrieval with an evolving structured representation of the reconstructed chain. Five complementary metrics are introduced to assess distinct stages of the reconstruction process and support systematic failure diagnosis. Based on evaluations using 6 LLMs, DiagChain reveals that even the strongest configuration succeeds on only 39.6\% of the 849 reference steps in MAIN-69. Our analysis further shows that smaller models struggle with the more basic task of incorporating retrieved evidence into their outputs, whereas larger models can proceed to later steps, where correctly ordering that evidence becomes the main bottleneck. These results validate the importance of diagnostic evaluation beyond end-to-end accuracy and provide actionable insights for improving evidence-grounded cybersecurity agents.
 
\end{abstract}\begin{center}{\small\textbf{Code and data:} \url{https://github.com/abrahaamm/DiagChain}}\end{center}

\section{Introduction}
Large Language Model (LLM) agents increasingly inspect logs, invoke tools, and reason over heterogeneous security telemetry during cyber investigations~\cite{Wu2025ExCyTInBench,Chona2026CyberDefenseBenchmark,SIRBench2026}. Recent benchmarks have expanded from security question answering and log analysis~\cite{CTIBench2024,Karlsen2024LLMLogAnalysis} to interactive threat hunting, incident triage, and forensic analysis~\cite{Wu2025ExCyTInBench,Chona2026CyberDefenseBenchmark,SIRBench2026,Anand2026AuditBench}. Meanwhile, studies of attack chains show that provenance and evidence are important to exhibit attack narratives or the reconstruction of ordered attack stages~\cite{Hossain2017SLEUTH,Cui2026SynthChain,Tan2026FuseChain}. These efforts demonstrate progress toward more realistic and evidence-aware security agents.

However, existing benchmarks still focus largely on final outputs or aggregate accuracy, offering limited visibility into the intermediate investigation process~\cite{CTIBench2024,Karlsen2024LLMLogAnalysis,Chona2026CyberDefenseBenchmark,Sun2026HIDBench}. Consequently, they provide only a partial view of an agent's ability to reconstruct an attack chain from observed evidence. We refer to this ability as \textit{Evidence-Grounded Attack Chain Reconstruction}. Existing benchmarks remain limited in evaluating it along three dimensions:
\begin{figure*}[!t]
\centering
\includegraphics[width=\textwidth]{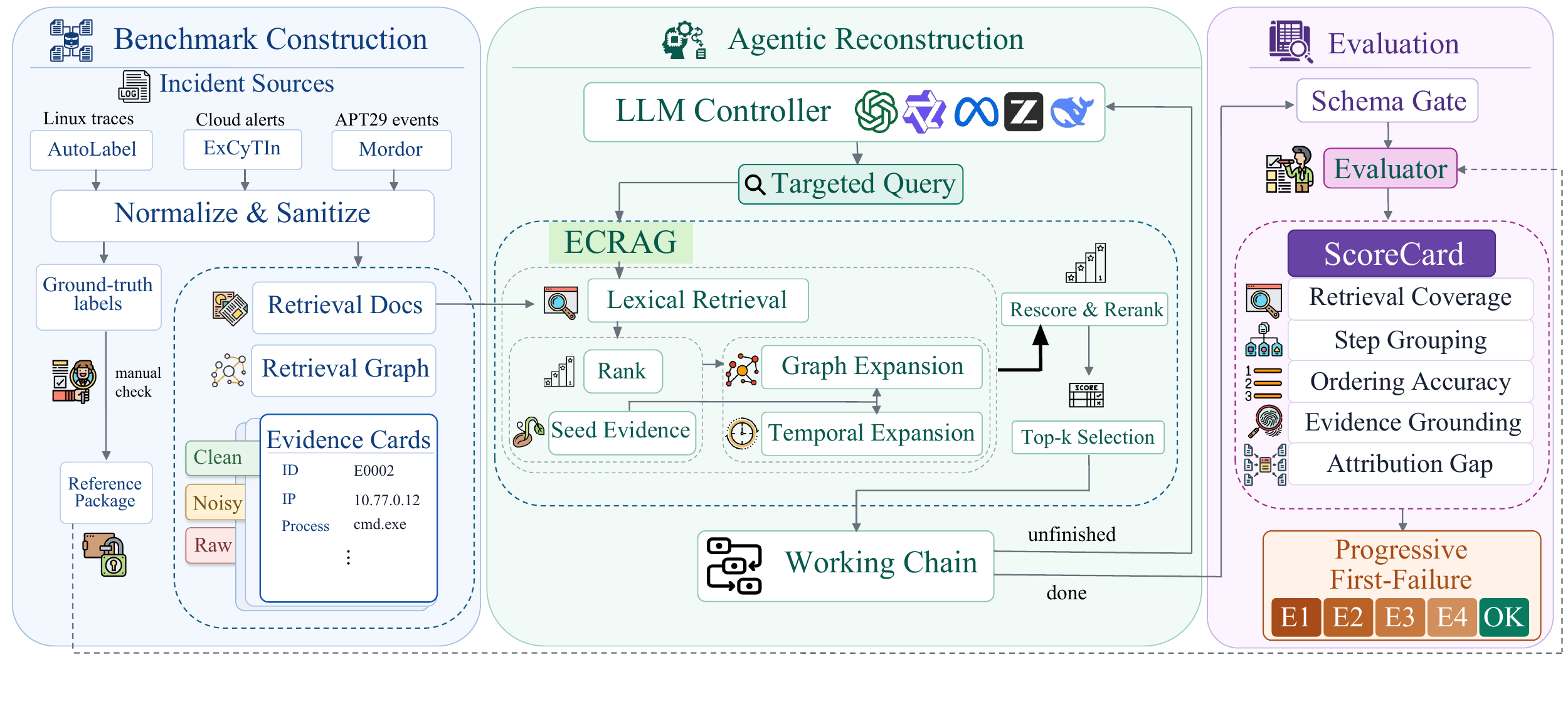}
\caption{Overview of the DiagChain benchmark, agentic reconstruction workflow, and diagnostic evaluation architecture.}
\label{fig:diagchain-overview}
\end{figure*}

\textbf{Limitation I: Scenario coverage.}
Existing datasets rarely span multiple systems, controlled noise levels, and diverse chain lengths~\cite{Wu2025ExCyTInBench,Anand2026AuditBench,Sun2026HIDBench,Cadet2026RAGSIA}.

\textbf{Limitation II: Diagnostic evaluation.}
Answers, reports, alerts, or aggregate chain scores do not reveal whether failure arose during evidence discovery, grouping, ordering, grounding, or attribution. Some benchmarks provide partial process-level diagnostics through intermediate-path rewards, tactic-level coverage, or attack-stage breakdowns, but they do not jointly localize failures across the complete reconstruction process~\cite{Wu2025ExCyTInBench,Chona2026CyberDefenseBenchmark,Cadet2026RAGSIA,Cui2026SynthChain}. 

\textbf{Limitation III: Investigation workflow.}
Tool-using investigation is commonly evaluated through interactive security tasks~\cite{Wu2025ExCyTInBench,Chona2026CyberDefenseBenchmark,SIRBench2026,Jajodia2026SIABench,Anand2026AuditBench}, whereas chain-generation studies typically assess ordered outputs without an evolving evidence-linked working state~\cite{Cui2026SynthChain,Tan2026FuseChain,Aly2025OCRAPT}.

To resolve the above limitations, we present \textbf{DiagChain} as a diagnostic benchmark for evidence-grounded attack chain reconstruction. We reconstruct open-source logs from Linux, Windows, enterprise, and cloud environments~\cite{Peng2025AutoLabel,Wu2025ExCyTInBench,OTRFSecurityDatasetsAPT29} into attack chain scenarios and organize them by three evidence noise levels and three chain length buckets, forming a representative dataset named MAIN-69. As for ground-truth labels, we adopt SynthChain's formulation of an attack chain as an ordered sequence of adversarial steps aligned with supporting telemetry events~\cite{Cui2026SynthChain}, and extend each step with participating entities and evidence identifiers. One author then manually checked all 69 reference packages against their source records, covering 849 steps and 780 temporal edges; no correction was required. Our task asks an agent to recover an ordered sequence of attacker actions, each grounded in evidence observed during a tool-using investigation~\cite{Yao2023ReAct,Wu2025ExCyTInBench,Chona2026CyberDefenseBenchmark}. Evidence-Centric Retrieval-Augmented Generation (ECRAG), building on retrieval-augmented generation~\cite{Lewis2020RAG}, supports this process by coupling evidence retrieval with an evolving structured representation of the reconstructed chain. The evaluator separately measures evidence discovery, grouping, ordering, grounding, and attribution, allowing model performance to be compared while locating failures within the reconstruction process.

We evaluate six models on MAIN-69, diagnose reconstruction failures by stage, and study the effects of agent scaffolding and retrieval budgets. The results reveal that smaller models more often fail to incorporate evidence they have already observed, whereas larger models struggle to order acquired evidence correctly. Even the strongest configuration completes only 39.6\% of reference steps without an earlier failure.
Overall, this paper makes four contributions:
\begin{itemize}
    \item \textbf{MAIN-69 scenarios.} We reconstruct open-source logs into 69 scenarios covering multiple systems and telemetry types, three noise profiles, and diverse chain lengths.
    \item \textbf{Diagnostic evaluation.} We introduce evidence-cluster alignment and five metrics that separate evidence discovery, grouping, ordering, grounding, and attribution. 
    \item \textbf{Agent workflow with ECRAG.} We couple Evidence-Centric Retrieval-Augmented Generation (ECRAG) with an evolving structured working chain, support checks, and grounded submission, making the conversion from evidence to chain explicit and auditable.
    \item \textbf{Benchmark and findings.} We benchmark six model configurations and expose distinct bottlenecks in evidence acquisition and chain assembly, including how chain length and evidence noise shift them.
\end{itemize}

\section{Related Work}

\begin{table*}[!t]
\centering
{\fontsize{9}{10.2}\selectfont
\setlength{\tabcolsep}{5.0pt}
\begin{tabular}{@{}lcccccc@{}}
\toprule
Work & Agent & RAG & Chain & Multi-sys. & Diag. & Len./Noise \\
\midrule
CTIBench \cite{CTIBench2024} & -- & -- & -- & -- & -- & -- \\
AttackSeqBench \cite{AttackSeqBench2025} & -- & Y & P & -- & -- & -- \\
ExCyTIn-Bench \cite{Wu2025ExCyTInBench} & Y & P & -- & -- & P & P \\
Cyber Defense Benchmark \cite{Chona2026CyberDefenseBenchmark} & Y & -- & -- & -- & P & -- \\
SIR-Bench \cite{SIRBench2026} & Y & -- & -- & -- & -- & -- \\
SIABench \cite{Jajodia2026SIABench} & Y & -- & -- & P & -- & -- \\
AuditBench \cite{Anand2026AuditBench} & -- & -- & -- & Y & -- & P \\
HIDBench \cite{Sun2026HIDBench} & -- & -- & Y & Y & -- & P \\
RAG-SIA \cite{Cadet2026RAGSIA} & -- & Y & Y & Y & Y & P \\
OCR-APT \cite{Aly2025OCRAPT} & -- & Y & Y & Y & Y & -- \\
SynthChain \cite{Cui2026SynthChain} & -- & -- & Y & Y & Y & -- \\
FuseChain \cite{Tan2026FuseChain} & -- & P & Y & Y & Y & -- \\
ProvSEEK \cite{Mukherjee2025ProvSEEK} & Y & Y & P & Y & -- & -- \\
\midrule
\textbf{DiagChain (ours)} & Y & Y & Y & Y & Y & Y \\
\bottomrule
\end{tabular}
}
\caption{Capabilities of representative related work. Chain, explicit ordered attack-chain output; Multi-sys., evidence from multiple operating-system or independently deployed system families; Diag., diagnosis at distinct steps or stages within the same attack chain; and Len./Noise, model performance stratified by chain length and evidence-noise level (Y=both; P=one axis or a related path-length, context-span, or noise/representation proxy with reported task-quality effects). Elsewhere, Y=explicit support, P=partial or indirect support, and --=not a primary evaluated target.}
\label{tab:related-work-capabilities}
\end{table*}

\subsection{LLM Agents for Cybersecurity}

General-purpose LLM agents interleave reasoning with actions and may retain verbal feedback in episodic memory \cite{Yao2023ReAct,Shinn2023Reflexion}; cybersecurity systems add log-query interfaces, forensic tools, and persistent investigation state. Interactive systems show that tool access supports more realistic threat hunting and triage, but also makes outcomes sensitive to action formulation, tool choice, memory, and budget \cite{Wu2025ExCyTInBench,Chona2026CyberDefenseBenchmark,Jajodia2026SIABench,SIRBench2026,Anand2026AuditBench}. Retrieval-augmented agents further integrate CTI, knowledge graphs, provenance, and forensic traces; comparisons consistently find that evidence selection and integration, rather than retrieval alone, determine downstream reliability \cite{Cadet2026RAGSIA,Hamzic2026BeyondRAGCTI,CyberRAG2025,Mukherjee2025ProvSEEK,Cheng2025OmniSec}.

\subsection{Attack-Chain Reconstruction}

Attack detection and investigation over endpoint telemetry have attracted sustained attention from both academia and industry \cite{xu2025deep,xu2026cerberus}. Attack-chain reconstruction evolved from alert correlation and attack-plan recognition to compact scenario recovery from system provenance \cite{Ning2002AttackScenarios,Qin2004AttackPlan,Hossain2017SLEUTH,Milajerdi2019HOLMES}. Provenance systems support CTI-to-audit graph alignment for threat hunting (POIROT), automated alert triage (NoDoze), and anomaly-based APT detection over long-running provenance (UNICORN) \cite{Milajerdi2019POIROT,Hassan2019NoDoze,Han2020UNICORN}; later graph methods improve scale and robustness to background activity \cite{Alsaheel2021ATLAS,Cheng2024Kairos,Jia2023MAGIC,Jiang2025ORTHRUS}. SLEUTH is particularly relevant because it reconstructs concise attack scenarios from audit-event provenance, which motivates DiagChain's evidence-linked chain model \cite{Hossain2017SLEUTH}. Recent LLM and multi-source studies extend reconstruction toward host-intrusion narratives, runtime attack chains, and interleaved APT behaviors \cite{Sun2026HIDBench,Cui2026SynthChain,Tan2026FuseChain,TGCM2026}.

\subsection{Benchmarks for Cybersecurity Investigation}

Cybersecurity benchmarks have moved from static knowledge and extraction tests toward interactive investigation. Static suites evaluate CTI knowledge and attack-sequence reasoning through question answering, whereas newer environments require models to query structured or raw telemetry, use tools, and operate under turn or cost constraints \cite{CTIBench2024,AttackSeqBench2025,Wu2025ExCyTInBench,Chona2026CyberDefenseBenchmark}. Their results show that success depends on investigation-path length, search scope, context representation, and budget, motivating evaluations of intermediate progress, evidence use, tool appropriateness, and failure type in addition to final accuracy \cite{Wu2025ExCyTInBench,Chona2026CyberDefenseBenchmark,SIRBench2026,Jajodia2026SIABench,Anand2026AuditBench}.

ExCyTIn-Bench is especially relevant because it derives questions and intermediate rewards from paths in alert-entity investigation graphs, then evaluates agents through SQL interaction; even strong models leave substantial headroom, and performance varies with path length and database scope \cite{Wu2025ExCyTInBench}. Other recent benchmarks expand evaluation toward open-ended threat hunting, replayed incident response, audit-log reasoning, and structured incident outputs \cite{Chona2026CyberDefenseBenchmark,SIRBench2026,Jajodia2026SIABench,Anand2026AuditBench,Sun2026HIDBench,Cadet2026RAGSIA,Cui2026SynthChain,Tan2026FuseChain,Aly2025OCRAPT}. Collectively, this literature shifts evaluation from what models know toward how they investigate, but it still primarily scores answers, flags, reports, or system-produced chains. Table~\ref{tab:related-work-capabilities} summarizes the methodological differences.

\section{DiagChain: Benchmark and System Design}

\subsection{Dataset Construction}

No single public source spans the system, telemetry, and chain-length variation needed for evaluating attack-chain reconstruction. We construct our dataset from 23 source units drawn from 13 selected AutoLabel scenarios~\cite{Peng2025AutoLabel}, all eight ExCyTIn-Bench incidents~\cite{Wu2025ExCyTInBench}, and the Day 1 and Day 2 host-event datasets from the APT29 in OTRF Security-Datasets~\cite{OTRFSecurityDatasetsAPT29}. Each source unit is transformed into clean, noisy, and raw evidence profiles, yielding 69 reconstruction cases, and we name the whole dataset MAIN-69. \emph{Clean} retains the core attack evidence, \emph{noisy} adds sampled benign or distracting activity, and \emph{raw} exposes a broader original or capped source window. Because the sources retain different platforms, evidence granularities, and native chain units, they are complementary rather than directly interchangeable; Appendix Table~\ref{tab:app-dataset-scale} reports the precise inclusion and processing scope. Table~\ref{tab:benchmark-composition} summarizes the resulting cases; S/M/L denote chains with 2--5, 6--15, and at least 16 reference steps.

\begin{center}
\centering
\small
\setlength{\tabcolsep}{3pt}
\begin{tabular*}{\columnwidth}{@{\extracolsep{\fill}}llcc@{}}
\toprule
Source ($cases$) & Native telemetry & Unit & S/M/L \\
\midrule
AutoLabel (39) & Linux mixed logs & Action & 33/6/0 \\
ExCyTIn (24) & Enterprise/cloud tables & Alert & 3/12/9 \\
OTRF APT29 (6) & Windows host events & Event & 0/0/6 \\
\bottomrule
\end{tabular*}
\captionof{table}{Composition of MAIN-69. ``($\cdot$)'' gives reconstructed-case totals; ``S/M/L'' gives their chain-length distribution.}
\label{tab:benchmark-composition}
\end{center}

Parsers normalize timestamps, source fields, and observable entities while removing answer-derived information. For each reconstruction case, the pipeline produces evidence cards, retrieval documents, an evidence--entity retrieval graph, and a gold reference package. The gold reference package is constructed by instantiating ordered reference steps, mapping their support to normalized evidence-card IDs, and deriving reference edges. We manually check every gold reference package before inclusion; Appendix~\ref{app:benchmark-construction} provides the complete procedure. Each reconstruction case contains the following artifacts:

\begin{itemize}
    \item \textbf{Evidence card.} The atomic evidence unit exposed to the agent. It stores an evidence ID, source provenance, type, an optional timestamp, observation content, and automatically extracted clues such as entities, paths, and so on.
    \item \textbf{Retrieval document.} A one-to-one searchable rendering of an evidence card. It serializes the card ID, sanitized source metadata, observation content, and extracted clues into text for lexical ranking, without adding evidence.
    \item \textbf{Evidence--entity retrieval graph.} An index over evidence cards, visible entities, and log sources. Its edges encode entity mentions, source membership, and local record adjacency for retrieval expansion.
    \item \textbf{Gold reference package.} The manually checked, model-hidden evaluation target. It contains a gold reference chain of ordered reference steps, their entities and supporting evidence IDs, reference edges, and case metadata; the agent never accesses it before evaluation.
\end{itemize}

\subsection{Agentic Reconstruction with ECRAG}

RAG combines parametric generation with retrieved non-parametric memory \cite{Lewis2020RAG}; ECRAG adapts this idea to evidence cards distributed across sources, where a single retrieval pass may be insufficient.

\textbf{Workflow and state.} The environment uses the artifacts constructed above: \(D\) is the searchable retrieval documents, \(E\) is the normalized evidence cards returned to the agent, and \(G\) is the evidence--entity retrieval graph, which supports ECRAG expansion by linking evidence cards to visible entities. At turn \(t\), state \(S_t=(O,C,m)\) comprises observed cards, the working chain, and compact memory. The model reads \(I\) and \(S_t\), chooses a typed action, and may revise \(C\); every step \(c_i\) cites evidence identifiers \(e_i\). Investigation and chain generation therefore update the same evolving state rather than forming separate stages (Algorithm~\ref{alg:agent-loop}).

\begin{table*}[t]
\centering
{\small
\setlength{\tabcolsep}{2.4pt}
\begin{tabular*}{\textwidth}{@{\extracolsep{\fill}}l*{18}{c}@{}}
\toprule
Metric & \multicolumn{3}{c}{\makebox[0pt][c]{Qwen-3-32b}} & \multicolumn{3}{c}{\makebox[0pt][c]{DeepSeek-V4-Pro}} & \multicolumn{3}{c}{\makebox[0pt][c]{GLM-5.2}} & \multicolumn{3}{c}{\makebox[0pt][c]{GLM-5.2-T}} & \multicolumn{3}{c}{\makebox[0pt][c]{Llama4-17b-Scout}} & \multicolumn{3}{c}{\makebox[0pt][c]{GPT-5.5}} \\
\cmidrule(lr){2-4} \cmidrule(lr){5-7} \cmidrule(lr){8-10} \cmidrule(lr){11-13} \cmidrule(lr){14-16} \cmidrule(lr){17-19}
\textit{(Clean)} & S & M & L & S & M & L & S & M & L & S & M & L & S & M & L & S & M & L \\
\midrule
Ret.$\uparrow$ & 1.00 & 0.98 & 0.86 & 1.00 & 0.75 & 0.89 & 1.00 & 0.92 & 0.78 & 1.00 & 1.00 & 0.97 & 1.00 & 1.00 & 0.91 & 1.00 & 1.00 & 0.94 \\
Grp.$\uparrow$ & 0.66 & 0.68 & 0.25 & 0.78 & 0.68 & 0.64 & 0.69 & 0.73 & 0.68 & 0.82 & 0.80 & 0.60 & 0.74 & 0.43 & 0.18 & 0.75 & 0.90 & 0.81 \\
Ord.$\uparrow$ & 0.90 & 0.97 & 0.73 & 1.00 & 0.95 & 0.84 & 1.00 & 0.97 & 0.82 & 1.00 & 0.99 & 0.85 & 0.97 & 0.72 & 0.60 & 1.00 & 0.96 & 0.90 \\
Grd.$\uparrow$ & 0.53 & 0.79 & 0.55 & 0.74 & 0.85 & 0.80 & 0.72 & 0.75 & 0.84 & 0.81 & 0.84 & 0.75 & 0.73 & 0.66 & 0.39 & 0.60 & 0.92 & 0.88 \\
Gap$\downarrow$ & 0.31 & 0.33 & 0.59 & 0.19 & 0.11 & 0.21 & 0.09 & 0.09 & 0.03 & 0.22 & 0.02 & 0.22 & 0.36 & 0.57 & 0.69 & 0.47 & 0.09 & 0.04 \\
\bottomrule
\end{tabular*}
\vspace{3pt}
\begin{tabular*}{\textwidth}{@{\extracolsep{\fill}}l*{18}{c}@{}}
\toprule
Metric & \multicolumn{3}{c}{\makebox[0pt][c]{Qwen-3-32b}} & \multicolumn{3}{c}{\makebox[0pt][c]{DeepSeek-V4-Pro}} & \multicolumn{3}{c}{\makebox[0pt][c]{GLM-5.2}} & \multicolumn{3}{c}{\makebox[0pt][c]{GLM-5.2-T}} & \multicolumn{3}{c}{\makebox[0pt][c]{Llama4-17b-Scout}} & \multicolumn{3}{c}{\makebox[0pt][c]{GPT-5.5}} \\
\cmidrule(lr){2-4} \cmidrule(lr){5-7} \cmidrule(lr){8-10} \cmidrule(lr){11-13} \cmidrule(lr){14-16} \cmidrule(lr){17-19}
\textit{(Noisy)} & S & M & L & S & M & L & S & M & L & S & M & L & S & M & L & S & M & L \\
\midrule
Ret.$\uparrow$ & 0.90 & 0.84 & 0.77 & 0.98 & 0.82 & 0.65 & 0.89 & 0.76 & 0.65 & 0.93 & 0.79 & 0.77 & 0.88 & 0.87 & 0.74 & 0.92 & 0.96 & 0.85 \\
Grp.$\uparrow$ & 0.64 & 0.57 & 0.21 & 0.62 & 0.54 & 0.47 & 0.59 & 0.55 & 0.52 & 0.64 & 0.52 & 0.54 & 0.72 & 0.47 & 0.21 & 0.66 & 0.49 & 0.59 \\
Ord.$\uparrow$ & 0.75 & 0.98 & 0.48 & 0.92 & 0.93 & 0.88 & 1.00 & 0.94 & 0.86 & 1.00 & 0.98 & 0.89 & 0.72 & 0.67 & 0.80 & 1.00 & 0.96 & 0.97 \\
Grd.$\uparrow$ & 0.39 & 0.54 & 0.34 & 0.54 & 0.57 & 0.71 & 0.64 & 0.59 & 0.68 & 0.70 & 0.50 & 0.72 & 0.24 & 0.51 & 0.51 & 0.63 & 0.59 & 0.73 \\
Gap$\downarrow$ & 0.58 & 0.17 & 0.73 & 0.42 & 0.26 & 0.34 & 0.19 & 0.03 & 0.01 & 0.22 & 0.18 & 0.09 & 0.61 & 0.63 & 0.65 & 0.40 & 0.11 & 0.15 \\
\bottomrule
\end{tabular*}
\vspace{3pt}
\begin{tabular*}{\textwidth}{@{\extracolsep{\fill}}l*{18}{c}@{}}
\toprule
Metric & \multicolumn{3}{c}{\makebox[0pt][c]{Qwen-3-32b}} & \multicolumn{3}{c}{\makebox[0pt][c]{DeepSeek-V4-Pro}} & \multicolumn{3}{c}{\makebox[0pt][c]{GLM-5.2}} & \multicolumn{3}{c}{\makebox[0pt][c]{GLM-5.2-T}} & \multicolumn{3}{c}{\makebox[0pt][c]{Llama4-17b-Scout}} & \multicolumn{3}{c}{\makebox[0pt][c]{GPT-5.5}} \\
\cmidrule(lr){2-4} \cmidrule(lr){5-7} \cmidrule(lr){8-10} \cmidrule(lr){11-13} \cmidrule(lr){14-16} \cmidrule(lr){17-19}
\textit{(Raw)} & S & M & L & S & M & L & S & M & L & S & M & L & S & M & L & S & M & L \\
\midrule
Ret.$\uparrow$ & 0.43 & 0.84 & 0.57 & 0.67 & 0.84 & 0.53 & 0.64 & 0.79 & 0.52 & 0.64 & 0.84 & 0.77 & 0.49 & 0.84 & 0.65 & 0.83 & 0.92 & 0.72 \\
Grp.$\uparrow$ & 0.67 & 0.54 & 0.25 & 0.68 & 0.56 & 0.52 & 0.61 & 0.54 & 0.59 & 0.59 & 0.52 & 0.59 & 0.76 & 0.45 & 0.20 & 0.57 & 0.48 & 0.53 \\
Ord.$\uparrow$ & 0.33 & 0.89 & 0.65 & 0.67 & 0.92 & 0.76 & 0.75 & 1.00 & 0.83 & 0.75 & 0.93 & 0.90 & 0.42 & 0.72 & 0.80 & 0.83 & 0.91 & 0.93 \\
Grd.$\uparrow$ & 0.16 & 0.51 & 0.51 & 0.44 & 0.64 & 0.58 & 0.48 & 0.52 & 0.68 & 0.49 & 0.54 & 0.66 & 0.25 & 0.51 & 0.42 & 0.52 & 0.54 & 0.77 \\
Gap$\downarrow$ & 0.40 & 0.58 & 0.82 & 0.25 & 0.26 & 0.10 & 0.16 & 0.08 & 0.05 & 0.19 & 0.09 & 0.10 & 0.40 & 0.54 & 0.66 & 0.39 & 0.07 & 0.25 \\
\bottomrule
\end{tabular*}
}
\caption{(RQ1) MAIN-69 case-macro averages by evidence profile, model, and standard chain length (S/M/L). GLM-5.2-T is the thinking-enabled run.}
\label{tab:rq1-main69-stratified}
\end{table*}

\begin{figure*}[t]
\centering
\includegraphics[width=\textwidth]{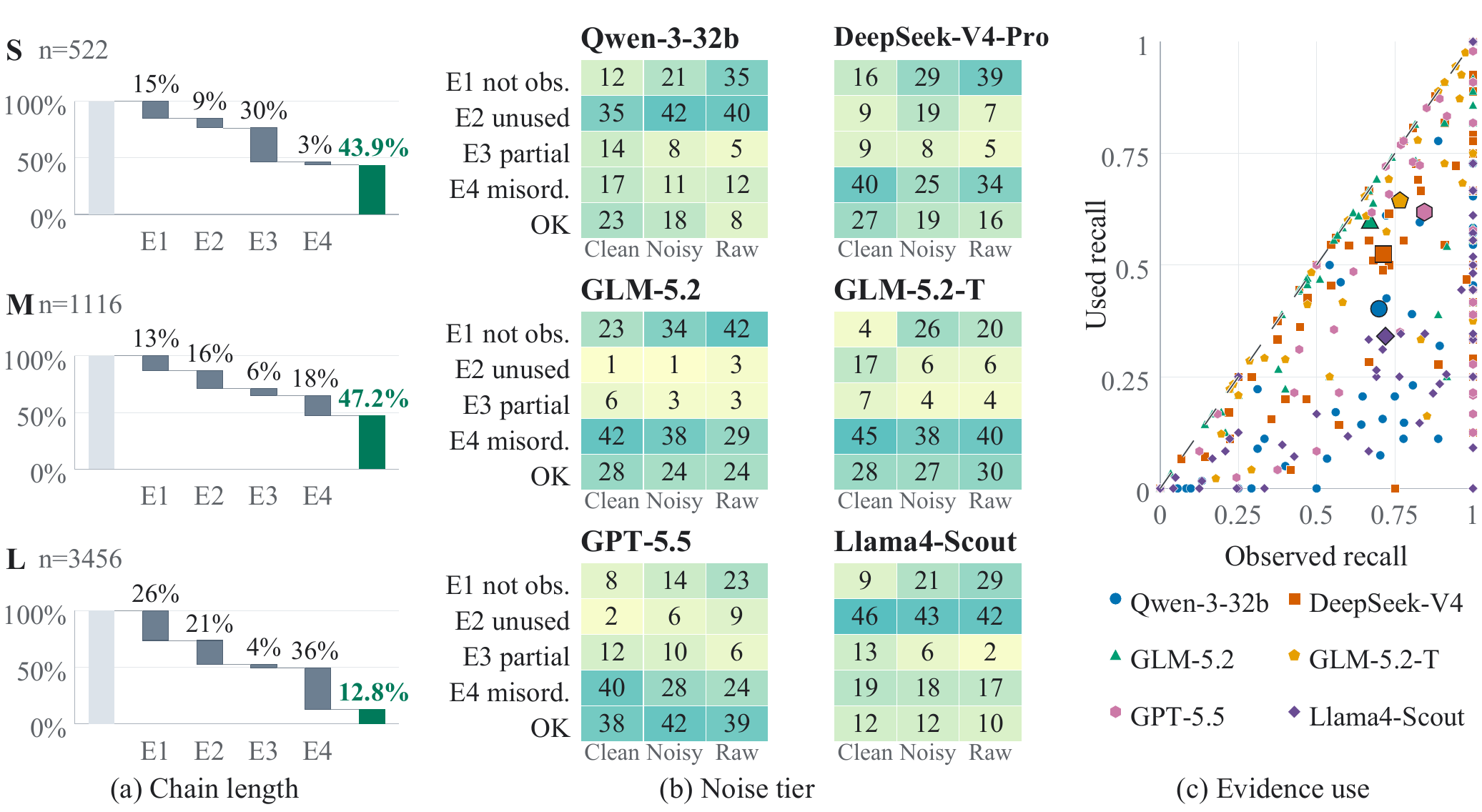}
\caption{(RQ2) diagnostics: (a) first failures by chain length; (b) first failures by model/profile; (c) observed ($x$) vs.\ step-cited ($y$) gold-evidence recall. In (c), faint/outlined markers are case runs/model means; dashed line: $y=x$.}
\label{fig:rq2-diagnostics-composite}
\end{figure*}

\begin{algorithm}[!htb]
\caption{Evidence-Grounded Agent Workflow}
\label{alg:agent-loop}
\begin{algorithmic}[1]
\REQUIRE Brief \(I\), environment \((D,E,G)\), model \(M\), budget \(T\)
\ENSURE Grounded chain \(Y\) and trace \(\tau\)
\STATE \(O,C,m,\tau\leftarrow\emptyset\)
\FOR{\(t=1,\ldots,T\)}
    \STATE \((a_t,q_t,\widetilde C)\leftarrow M(I,O,C,m)\)
    \STATE \(C\leftarrow\textsc{Validate}(\widetilde C)\)
    \IF{\(a_t=\texttt{submit}\)}
        \STATE \(U\leftarrow\{c_i\in C:e_i\cap\textsc{Ids}(O)=\emptyset\}\)
        \IF{\(U=\emptyset\) or \(t=T\)} \STATE \textbf{break} \ENDIF
        \STATE \(a_t\leftarrow\texttt{query\_evidence}\)
        \STATE \(q_t\leftarrow\langle\mbox{claims, entities, and times in }U\rangle\)
    \ENDIF
    \IF{\(a_t=\texttt{query\_evidence}\)}
        \STATE \(o_t\leftarrow\textsc{ECRAG}(q_t,D,E,G,k)\)
    \ELSE
        \STATE \(o_t\leftarrow\textsc{Execute}(a_t,q_t,O,C,E)\)
    \ENDIF
    \STATE \(O\leftarrow O\cup o_t;\quad m\leftarrow\textsc{UpdateMemory}(m,o_t,C)\)
    \STATE Append \((a_t,q_t,o_t)\) to \(\tau\)
\ENDFOR
\STATE \(\widehat C\leftarrow M(I,O,C;\textit{revise})\)
\STATE \(e_i\leftarrow e_i\cap\textsc{Ids}(O),\quad \forall c_i\in\widehat C\)
\STATE \(Y\leftarrow\widehat C[\{c_i:e_i\neq\emptyset\}]\)
\STATE \textbf{return} \(Y,\tau\)
\end{algorithmic}
\end{algorithm}

Within the loop, \texttt{query\_evidence} invokes ECRAG; other typed actions inspect observed cards or check source-local temporal order. Unsupported steps trigger further retrieval, and finalization retains only observed citations and supported steps. Exact budgets, retrieval coefficients, stop/repair rules, model requests, and prompts appear in Appendix~\ref{app:agent-ecrag-config}.

\textbf{Evidence retrieval with ECRAG.} ECRAG starts from a simple problem: a useful log record may not repeat the words in the agent's query. It may instead mention the same host, account, process, or file as a matched record, or appear immediately before or after that record. Keyword search alone can therefore miss evidence needed for the next chain step.

ECRAG retrieves evidence in three steps. First, it ranks retrieval documents with TF-IDF~\cite{SaltonBuckley1988} and keeps a small set of strong textual matches as seeds. Second, it applies two parallel operations: Entity Expansion uses \(G\) to retrieve cards that share visible entities with a seed, whereas Temporal Expansion retrieves the immediately preceding and following records within the same source. Expansion stops after this local neighborhood. Third, ECRAG merges the candidates, ranks them again, and returns the top \(k\) cards in source-local record order.

\begin{algorithm}[!htb]
\caption{Evidence-Centric Retrieval (ECRAG)}
\label{alg:ecrag}
\begin{algorithmic}[1]
\REQUIRE Intent \(q\), artifacts \((D,E,G)\), budget \(k\)
\ENSURE Evidence observation \(o\)
\STATE \(R_0\leftarrow\textsc{Top}_{\max(k,8)}(\textsc{TFIDFRank}(q,D))\)
\STATE \(S\leftarrow\textsc{First}_{\min(8,|R_0|)}(R_0)\)
\STATE \(X_e\leftarrow\textsc{SharedEntityCards}(S,G)\)
\STATE \(X_t\leftarrow\textsc{AdjacentCards}(S,E,1)\)
\STATE \(X\leftarrow R_0\cup X_e\cup X_t\)
\STATE \(r(e)\leftarrow s_{\rm tfidf}(q,e)+s_{\rm tok}(q,e)\)
\STATE \(\hphantom{r(e)\leftarrow{}}+b_{\rm ent}(q,e)+b_{\rm meta}(e),\quad e\in X\)
\STATE \(K\leftarrow\textsc{Top}_{k}(X;r)\)
\STATE \(o\leftarrow\textsc{Sort}(E[K];\textit{source},\textit{local record})\)
\STATE \textbf{return} \(o\)
\end{algorithmic}
\end{algorithm}

The second ranking uses four simple signals. \(s_{\rm tfidf}\) measures textual similarity to the query, \(s_{\rm tok}\) measures token overlap, \(b_{\rm ent}\) rewards an exact visible-clue match, and \(b_{\rm meta}\) adds a bounded bonus from visible source and security fields. Appendix~\ref{app:agent-ecrag-config} gives the exact candidate limits, coefficients, and tie-breaking rules.

\subsection{Diagnostic Evaluator}

A final-answer score cannot distinguish failure to find evidence from failure to organize evidence already found. DiagChain therefore validates the schema, represents each step by its evidence cluster, and aligns predicted and reference clusters by maximum evidence overlap. 

\textbf{Retrieval Step Coverage (Ret.)} represents \textit{how much of the reference chain became available to the agent during investigation}. It is calculated as the fraction of reference steps for which the agent observed at least one supporting evidence item.

\textbf{Grouping F1 (Grp.)} represents \textit{how well supporting evidence is partitioned into coherent steps despite valid differences in chain length or granularity}. It is calculated by applying $B^3$~\cite{BaggaBaldwin1998} to the predicted and reference evidence-to-step assignments. It captures merge/split errors independently of retrieval.

\textbf{Ordering Accuracy (Ord.)} represents \textit{the fraction of comparable step pairs placed in the same order as the reference chain}. It is calculated as pairwise order agreement among aligned steps retained after matching, thereby scoring chronology independently of absolute indices and unmatched content.

\textbf{Evidence Grounding F1 (Grd.)} represents \textit{the precision and completeness of cited support within those aligned pairs}. It is calculated as the evidence-set F1 between each retained aligned predicted and reference step. Because unmatched predicted steps are not scored by Grd., it should be interpreted jointly with Grp., Ret., and the extra-step diagnostics reported in the appendix.

\textbf{Attribution Gap Rate (Gap)} represents \textit{how much relevant evidence the agent found but failed to use in its final attribution}. It is calculated as the fraction of observed reference evidence omitted from final step and causal-edge citations. It separates successful discovery from losses during evidence selection and chain assembly; lower is better.

Formal definitions and rules appear in Appendix~\ref{app:metrics-funnel}.

\section{Experiments}

This study aims to answer the following 4 research questions:

\textbf{RQ1 (Overall Capability):} How well do current LLMs perform on attack chain reconstruction?

\textbf{RQ2 (Failure Anatomy):} Where does reconstruction first fail?

\textbf{RQ3 (Scaffold Ablation):} What do retrieval and the full Agent workflow add?

\textbf{RQ4 (Budget Sensitivity):} How do retrieval depth and interaction turns affect quality and cost?

\noindent\textbf{Setup.}
We use MAIN-69 for RQ1/RQ2. Under a bounded evaluation budget, RQ3 uses R12, a 12-case subset selected from R24, while RQ4 uses R24, a 24-case subset of MAIN-69. Both diagnostic subsets cover the three sources, clean/noisy/raw evidence, and S/M/L chains. Unless otherwise stated, every condition uses the same prompt, a 15-turn ceiling, and retrieval width $k=32$. Temperature 0 is sent for the Ollama, DeepSeek, and GLM backends; the GPT-5.5 Responses request omits that field and uses the provider default.

\subsection{Main Results (RQ1)}

RQ1 asks how well current models do the task under the fixed setup above. Table~\ref{tab:rq1-main69-stratified} reports the stratified results.

No model dominates every reconstruction stage. Across the stratified cells in Table~\ref{tab:rq1-main69-stratified}, GPT-5.5 is most consistently strong in evidence retrieval and ordering, whereas GLM-5.2 often leaves the smallest attribution gap. The table also shows why retrieval alone is insufficient. On raw medium chains, Qwen-3-32b and DeepSeek-V4-Pro both obtain 0.84 retrieval coverage, yet DeepSeek reaches 0.64 grounding with a 0.26 gap, versus 0.51 and 0.58 for Qwen. 

The sharpest separation appears on long chains: Qwen-3-32b and Llama4-17b-Scout have grouping F1 values of only 0.18--0.25 across evidence profiles despite retrieval coverage of 0.57--0.91, while GPT-5.5 and GLM-5.2-Thinking maintain grouping F1 of 0.53--0.81 and 0.54--0.60, respectively. Model quality therefore depends on converting retrieved evidence into coherent, ordered, and grounded steps, not merely finding more evidence.

\subsection{Failure Anatomy (RQ2)}
RQ2 asks where the task first fails at the step level. Each coverable reference step is assigned to its first failed stage: evidence not observed (E1), observed but unused (E2), partially attributed (E3), misordered (E4), or correct reconstruction. Figure~\ref{fig:rq2-by-model} shows the results of six models.

\begin{center}
\centering
\includegraphics[width=0.97\columnwidth]{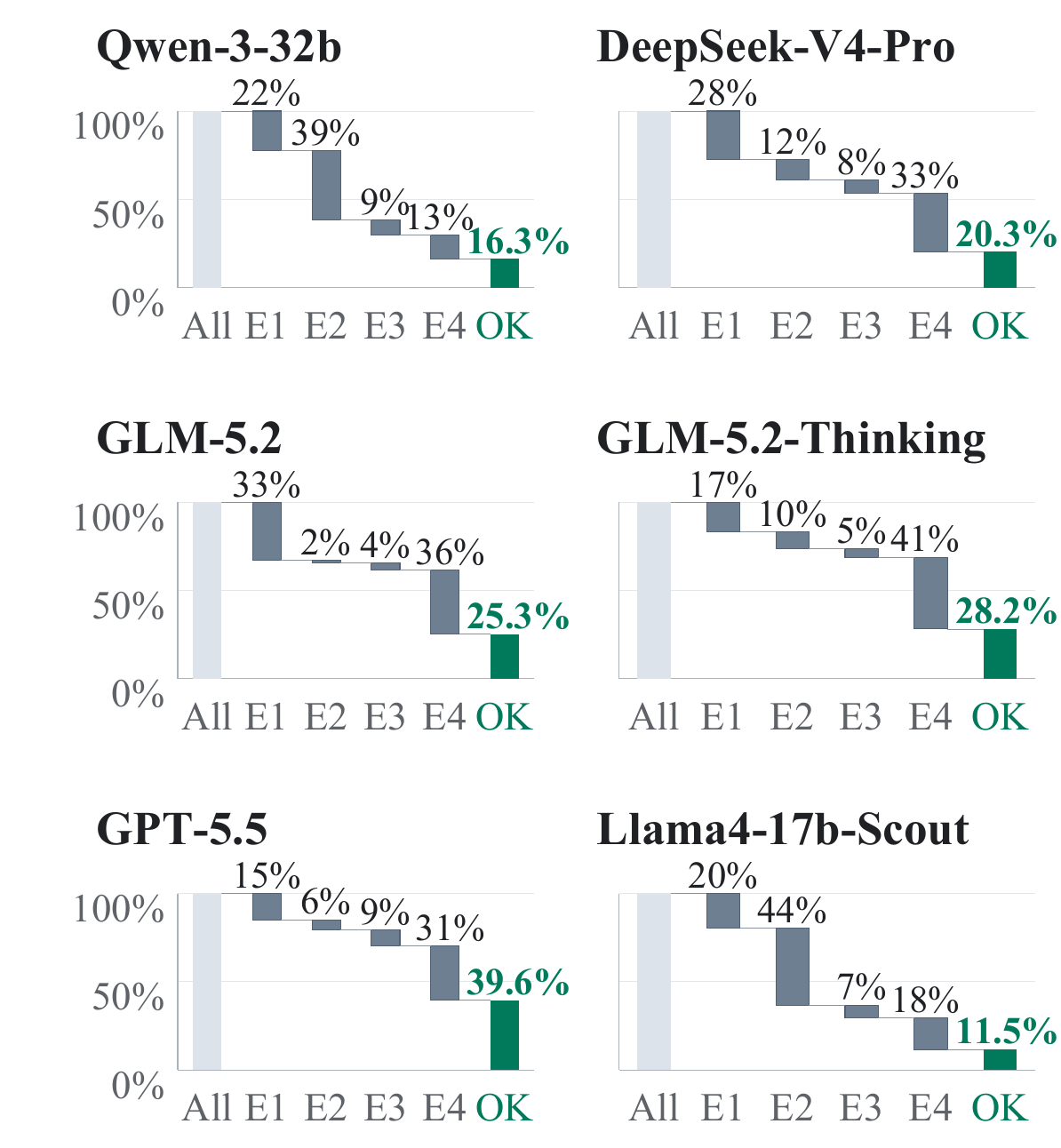}
\captionsetup{width=0.97\columnwidth}
\captionof{figure}{(RQ2) first-failure decomposition by model.}
\label{fig:rq2-by-model}
\end{center}

Results show larger models move the bottleneck downstream rather than eliminating it (Figure~\ref{fig:rq2-by-model}). Qwen-3-32b and Llama4-17b-Scout, the smaller configurations in our comparison, are dominated by observed but unused evidence, whereas DeepSeek-V4-Pro, GLM-5.2 variants, and GPT-5.5 with larger capacity or reasoning mode more often fail while ordering evidence already acquired. This split suggests a shift from evidence retention to global chain organization. Better evidence access therefore changes the form of failure, but does not by itself solve chain assembly.

Chain length and evidence noise are associated with different failure profiles. In this benchmark mixture, long chains show more ordering and assembly failures, combining missing, unused, and misordered evidence rather than producing one dominant error (Figure~\ref{fig:rq2-diagnostics-composite}(a)). Raw evidence is associated primarily with more evidence-not-observed failures across models (Figure~\ref{fig:rq2-diagnostics-composite}(b)). Because source family and step unit are correlated with chain length, these patterns are diagnostic rather than source-independent causal effects.

Figure~\ref{fig:rq2-diagnostics-composite}(c) separates evidence discovery from evidence use. All six model means lie below the diagonal. Llama4-17b-Scout shows the widest separation (0.720 observed versus 0.341 step-cited), followed by Qwen-3-32b (0.699 observed versus 0.402 step-cited). GPT-5.5 lies farthest to the right (0.845) but step-cites only 0.619, whereas GLM-5.2 observes less (0.671) yet step-cites 0.598 and lies closest to the diagonal. Thus, retrieval breadth and evidence conversion are distinct: finding more relevant evidence does not ensure that it is retained and attributed in the final chain. In this comparison, the smaller Qwen-3-32b and Llama4-17b-Scout means fall farther below the diagonal than the larger or reasoning-enhanced configurations, indicating weaker conversion of observed evidence into cited chain steps.

\subsection{Scaffold Ablation (RQ3)}

RQ3 evaluates three GLM-5.2 configurations on the same R12 cases (36 model--case runs). LLM only tests generation without retrieval; Retrieval only adds evidence access; and Full scaffold adds structured chain memory, typed investigation, reflection, support audit, and grounding checks. Because these post-retrieval mechanisms share the evolving chain state, we evaluate them as one reconstruction layer rather than assign causal credit to individual components.

Table~\ref{tab:rq3-glm52-ablation} shows that, on R12, retrieval alone is insufficient: it improves evidence access and omission, while Full scaffold trades small losses in retrieval coverage and grouping for much stronger ordering, grounding, and attribution. Under the fixed interaction budget, this trade-off is consistent with revision and support checks consuming capacity while rejecting weak assignments. Full scaffold therefore offers the best observed task-level trade-off on this diagnostic subset for converting found evidence into a coherent, grounded, and auditable chain.

\begin{table}[!htb]
\centering
\small
\setlength{\tabcolsep}{2.6pt}
\begin{tabular*}{\columnwidth}{@{\extracolsep{\fill}}lccccc@{}}
\toprule
Scaffold & Ret.$\uparrow$ & Grp.$\uparrow$ & Ord.$\uparrow$ & Grd.$\uparrow$ & Gap$\downarrow$ \\
\midrule
LLM only & 0.824 & 0.578 & 0.784 & 0.559 & 0.402 \\
Retrieval only & 0.863 & 0.565 & 0.764 & 0.573 & 0.306 \\
\midrule
Full scaffold & 0.817 & 0.519 & 0.943 & 0.610 & 0.054 \\
\bottomrule
\end{tabular*}
\caption{(RQ3) GLM-5.2 system ablation on R12. LLM only excludes evidence retrieval, Retrieval only adds the retrieval kernel, and Full scaffold evaluates the remaining reconstruction mechanisms as one integrated workflow. Values are case-macro averages.}
\label{tab:rq3-glm52-ablation}
\end{table}

\subsection{Budget Sensitivity (RQ4)}

\begin{figure}[!t]
\centering
\includegraphics[width=\linewidth]{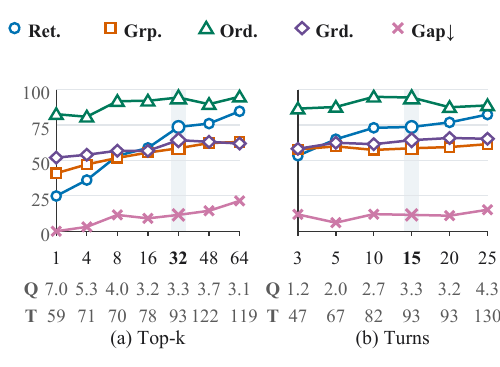}
\caption{(RQ4) budget sensitivity for GLM-5.2 on R24. Left: top-$k$ at 15 turns; right: turns at $k=32$. Shading marks the selected setting; Q and T report mean queries and tokens (thousands).}
\label{fig:rq4-budget-sensitivity}
\end{figure}

RQ4 tests whether a larger budget alone repairs reconstruction failures. Holding GLM-5.2 and the full scaffold fixed on R24, increasing $k$ or the turn ceiling expands evidence exposure, but the benefit does not consistently reach the final chain (Figure~\ref{fig:rq4-budget-sensitivity}). The low gap at $k=1$ is misleading because little reference evidence is observed. At $k=32$, the agent combines high coverage with peak grounding and stable ordering; beyond it, gains concentrate in coverage and grouping while unused evidence and cost grow. Likewise, turns beyond 15 raise coverage but destabilize ordering and do not reduce attribution errors, while mean token use rises from 93K at 15 turns to 130K at 25. We therefore use $k=32$ and 15 turns as a common operating point: enough headroom for retrieval, revision, and support checks, with early stopping limiting unnecessary work. This is an empirical default rather than a universal optimum; Appendix~\ref{app:additional-results} reports the full sweeps.

\section{Conclusion}

In this paper, we create DiagChain, a diagnostic benchmark for evidence-grounded attack chain reconstruction. It combines MAIN-69, which spans multiple systems, noise levels, and chain lengths; an ECRAG agent with a structured working chain; and five stage-specific metrics for task evaluation. Across six LLMs, no model dominates every reconstruction stage. Smaller models often lose observed evidence before submission, whereas larger or reasoning-enhanced models more often fail at global ordering. Raw evidence impairs discovery, while long chains expose assembly limits. DiagChain therefore provides an auditable testbed for cybersecurity agents. Beyond benchmarking, our findings show that stronger models shift rather than eliminate failures and that larger interaction budgets provide no consistent remedy, motivating diagnostic evaluation and evidence-aware agent design for reliable cybersecurity investigation.

\clearpage
\setcounter{table}{4}
\setcounter{figure}{6}
\appendix
\setcounter{secnumdepth}{1}
\makeatletter
\renewcommand{\@seccntformat}[1]{\csname the#1\endcsname.\quad}
\makeatother

\definecolor{appBlueFill}{HTML}{EAF2FA}
\definecolor{appBlueRule}{HTML}{3F6F9F}
\definecolor{appGreenFill}{HTML}{EAF5EF}
\definecolor{appGreenRule}{HTML}{2F7659}
\definecolor{appAmberFill}{HTML}{FFF3D6}
\definecolor{appAmberRule}{HTML}{946200}
\definecolor{appRoseFill}{HTML}{FBECEF}
\definecolor{appRoseRule}{HTML}{98485A}
\definecolor{appVioletFill}{HTML}{F1ECFA}
\definecolor{appVioletRule}{HTML}{6B4C92}
\definecolor{appGrayFill}{HTML}{F2F4F6}
\definecolor{appGrayRule}{HTML}{66717E}

\newcommand{\appcolpanel}[3]{%
  \noindent\fcolorbox{#1}{#2}{%
    \parbox{\dimexpr\columnwidth-2\fboxsep-2\fboxrule\relax}{%
      \vspace*{\dimexpr6pt-\fboxsep\relax}%
      \noindent\hspace*{\dimexpr6pt-\fboxsep\relax}%
      \parbox{\dimexpr\columnwidth-12pt-2\fboxrule\relax}{#3}%
      \par\vspace*{\dimexpr6pt-\fboxsep\relax}}}\par
}
\newsavebox{\apppaddedbox}
\newcommand{\appframewithpadding}[4]{%
  \begingroup
  \sbox{\apppaddedbox}{#4}%
  \fcolorbox{#1}{#2}{%
    \raisebox{0pt}%
      [\dimexpr\ht\apppaddedbox+#3-\fboxsep\relax]%
      [\dimexpr\dp\apppaddedbox+#3-\fboxsep\relax]{%
        \hspace*{\dimexpr#3-\fboxsep\relax}%
        \usebox{\apppaddedbox}%
        \hspace*{\dimexpr#3-\fboxsep\relax}}}%
  \endgroup
}

\newcommand{\appnoisecard}[3]{\begingroup\setlength{\fboxsep}{4pt}\fcolorbox{#1}{#2}{\parbox[t][2.34in][t]{0.285\textwidth}{#3}}\endgroup}
\newcommand{\appcasecard}[3]{%
  \appframewithpadding{#1}{#2}{5pt}{%
    \parbox[t][1.18in][t]{0.435\textwidth}{#3}}%
}

\lstdefinestyle{appprompt}{%
  basicstyle=\fontsize{9}{9.2}\selectfont\ttfamily,
  numbers=none,
  frame=single,
  frameround=tttt,
  rulecolor=\color{appBlueRule},
  backgroundcolor=\color{appGrayFill},
  framesep=3pt,
  xleftmargin=0pt,
  xrightmargin=0pt,
  columns=fullflexible,
  keepspaces=true,
  breaklines=true,
  breakatwhitespace=false,
  showstringspaces=false,
  aboveskip=0pt,
  belowskip=0pt}

\setcounter{dbltopnumber}{3}
\setcounter{topnumber}{3}
\renewcommand{\dbltopfraction}{0.99}
\renewcommand{\topfraction}{0.99}
\renewcommand{\textfraction}{0.01}
\renewcommand{\floatpagefraction}{0.95}

\section*{Appendix}

\section{LLM Usage}
\label{app:llm-usage}

LLM agents are the systems evaluated in this study, and their experimental use is described in the methodology. Separately, generative AI tools were used for grammar check and essential language polish. 

\section{Benchmark Construction and Leakage Boundary}
\label{app:benchmark-construction}

\newcommand{\appcmark}{\ensuremath{\surd}}

This section records only construction details omitted from the main paper: source-specific bounds and reference rules, and the audits applied before freezing the benchmark.

\subsection{Source Scope, Profiles, and Frozen Manifest}

Table~\ref{tab:app-dataset-scale} reports the retained benchmark scope for AutoLabel, ExCyTIn-Bench, and OTRF APT29. Counts describe the derived benchmark rather than native dataset sizes; the accompanying SHA-256 manifest uniquely identifies the exact paper inputs.

\begin{table*}[!t]
\centering
\small
\setlength{\tabcolsep}{3.5pt}
\begin{tabular}{@{}llrrrrrrl@{}}
\toprule
Source & Inclusion & Units & Cases & Ref. steps & Cards & Span & Graph nodes & Unit \\
\midrule
AutoLabel & Selected scenarios & 13 & 39 & 141 & 4,512 & 2--13 & 7,453 & Action \\
ExCyTIn-Bench (SecRL) & All incidents & 8 & 24 & 558 & 4,816 & 4--68 & 6,703 & Alert \\
\shortstack[l]{OTRF Security-Datasets: \\ APT29} & Both days & 2 & 6 & 150 & 3,505 & 24--26 & 3,607 & Event \\
\midrule
Total & -- & 23 & 69 & 849 & 12,833 & 2--68 & 17,763 & Heterogeneous \\
\bottomrule
\end{tabular}
\caption{Inclusion scope and scale of MAIN-69. Units denote native source scenarios, incidents, or attack days, whereas cases denote reconstructed variants after applying three evidence profiles. Thus, 39, 24, and 6 are benchmark-case counts rather than native dataset instance counts. Step spans are per benchmark case.}
\label{tab:app-dataset-scale}
\end{table*}

\begin{table*}[!t]
\centering
\small
\setlength{\tabcolsep}{3.2pt}
\begin{tabular}{@{}lp{0.23\textwidth}p{0.31\textwidth}p{0.29\textwidth}@{}}
\toprule
Source & Retained evidence fields & Profile construction & Frozen bounds \\
\midrule
AutoLabel & Event time/type/category; process, file, network, registry, and selected raw-result fields & Clean keeps support cards; noisy adds two background windows outside a 40-record guard band; raw keeps an 80-record window around support & Pre-deduplication background-to-support target 3:1; noisy at most 180 cards; raw at most 260 cards; each card at most 900 characters \\
ExCyTIn-Bench & Alert time, name, severity, description, provider/product, status, compromised entity, graph entities, and allowlisted security-table columns & Alert-only uses alert graphs; alert-evidence adds SecurityIncident, AlertInfo, and AlertEvidence rows; raw-sampled adds keyword/entity-matched rows from the wider table set & Alert-only at most 260 cards; alert-evidence 360 total/240 CSV/140 per table; raw 520 total/420 CSV/80 per table \\
OTRF APT29 & Windows event time/ID/channel/host; account, process, command, file, registry, and network fields & Clean keeps aligned support events; noisy adds chronological background; raw uses a 25-record window around support and is evenly subsampled if capped & At most four support events per step; noisy background at most 600 cards; raw at most 1,500 cards \\
\bottomrule
\end{tabular}
\caption{Source-specific parsing, evidence-profile construction, and sampling limits. All profiles remove answer markers, gold reference packages, attack scripts, scene configuration, and other model-hidden fields before card generation.}
\label{tab:app-parser-sampling}
\end{table*}

In the order of clean/noisy/raw, the profile names are \texttt{evidence-only}/\texttt{mixed-log}/\texttt{raw-log} for AutoLabel, \texttt{alert-only}/\texttt{alert-evidence}/\texttt{raw-sampled} for ExCyTIn-Bench, and \texttt{L-clean}/\texttt{L-noisy}/\texttt{L-raw} for OTRF APT29. Table~\ref{tab:app-parser-sampling} gives the retained fields and exact caps. Figure~\ref{fig:app-profile-excerpts} compares model-visible excerpts from one real AutoLabel case across the three profiles. \begin{figure*}[!t] \centering {\setlength{\fboxsep}{4pt}\fcolorbox{appGreenRule}{appGreenFill}{\parbox[t][2.34in][t]{0.285\textwidth}{\raggedright\normalfont\textbf{Clean (\texttt{evidence-only}; 2 cards)}\par\smallskip{\fontsize{6.7}{7.7}\selectfont\ttfamily [E0001] 03:34:12.347810\par \textcolor{appRoseRule}{\bfseries[ATTACK] GET /geoserver/ows}\par \textcolor{appRoseRule}{\bfseries CQL\_FILTER=\ldots SELECT version()\ldots}\par src=192.168.123.2\par dst=192.168.123.3:8080\par [E0002] 03:34:12.347811\par \textcolor{appRoseRule}{\bfseries[ATTACK] mirrored frame 6075}\par method=GET; host=192.168.123.3:8080\par service=wfs; request=GetFeature}}}}\hfill {\setlength{\fboxsep}{4pt}\fcolorbox{appAmberRule}{appAmberFill}{\parbox[t][2.34in][t]{0.285\textwidth}{\raggedright\normalfont\textbf{Noisy (\texttt{mixed-log}; 14 cards)}\par\smallskip{\fontsize{6.7}{7.7}\selectfont\ttfamily [E0001] GET /geoserver/web/\par [E0002] TCP 56642 -> 8080\par [E0003] GET \ldots MapPreviewPage\par [E0004] GET /geoserver/web/?320\par [E0005] GET /geoserver/web/?344\par [E0006] TCP 56642 -> 8080\par \textcolor{appRoseRule}{\bfseries[E0007] [ATTACK] GET /geoserver/ows}\par \textcolor{appRoseRule}{\bfseries CQL\_FILTER=\ldots SELECT version()\ldots}\par \textcolor{appRoseRule}{\bfseries[E0008] mirrored attack capture}\par [E0009] TCP 56642 -> 8080\par [E0010] GET /geoserver/web/?410\par [E0012] GET \ldots GetCapabilities\par [E0013] GET /geoserver/web\par \ldots\ 2 omitted (TCP): E0011, E0014}}}}\hfill {\setlength{\fboxsep}{4pt}\fcolorbox{appBlueRule}{appBlueFill}{\parbox[t][2.34in][t]{0.285\textwidth}{\raggedright\normalfont\textbf{Raw (\texttt{raw-log}; 258 cards)}\par\smallskip{\fontsize{6.7}{7.7}\selectfont\ttfamily \ldots\ earlier cards omitted\par [E0123] GET \ldots AboutGeoServerPage\par [E0124] TCP 56642 -> 8080\par [E0125] TCP 56642 -> 8080\par [E0126] TCP 46124 -> 8080\par [E0128] TCP 46124 -> 8080\par [E0129] TCP 46124 -> 8080\par \textcolor{appRoseRule}{\bfseries[E0130] [ATTACK] GET /geoserver/ows}\par \textcolor{appRoseRule}{\bfseries CQL\_FILTER=\ldots SELECT version()\ldots}\par \textcolor{appRoseRule}{\bfseries[E0131] mirrored attack capture}\par [E0132] GET \ldots AboutGeoServerPage\par [E0133] GET \ldots AboutGeoServerPage\par [E0134] TCP 56642 -> 8080\par [E0135] TCP 56642 -> 8080\par [E0136] GET /geoserver/web\par [E0137] GET /geoserver/web\par [E0138] GET /geoserver/web/\par \ldots\ later cards omitted}}}} \caption{Model-visible excerpts from the same real AutoLabel case (CVE-2023-25157) across the three evidence profiles. Bold rose text and the \texttt{[ATTACK]} tag identify the SQL-injection request; URI fragments are shortened and percent-decoded for legibility; ellipses mark omitted fields or cards. Clean is shown completely, whereas noisy and raw are deliberately truncated. Evidence IDs are profile-local.} \label{fig:app-profile-excerpts} \end{figure*}

Table~\ref{tab:app-reference-example} gives a GeoServer example with case-local support IDs; its attack-script anchors serve annotation audit only.

\begin{table*}[!t]
\centering
\small
\setlength{\tabcolsep}{4pt}
\begin{tabular}{lp{0.31\textwidth}lll}
\toprule
Artifact & Semantic content & Clean & Noisy & Raw \\
\midrule
G01 & Send the crafted GeoServer OWS request carrying the \texttt{CQL\_FILTER} SQL expression & E0001--E0002 & E0007--E0008 & E0130--E0131 \\
G02 & Trigger backend evaluation of \texttt{SELECT version()} through the vulnerable query path & E0001--E0002 & E0007--E0008 & E0130--E0131 \\
Edge & G01 $\rightarrow$ G02, \texttt{temporal\_after}; support derived conservatively from the target step & E0001--E0002 & E0007--E0008 & E0130--E0131 \\
Hidden provenance & Attack-script line anchors retained for annotation audit only & \multicolumn{3}{l}{not present in any incident input} \\
\bottomrule
\end{tabular}
\caption{Gold reference-package construction for the GeoServer example. The supporting evidence IDs differ across evidence profiles because each reconstruction case has its own case-local namespace.}
\label{tab:app-reference-example}
\end{table*}

Beyond the totals in Table~\ref{tab:app-dataset-scale}, the frozen manifest contains 780 reference edges, 1,533 support-cited cards, and 73,246 retrieval-graph edges. The other 11,300 cards are uncited, not necessarily benign, because raw windows may contain relevant but unlabeled telemetry.

\subsection{Reference Construction and Verification}

Reference candidates retain their source provenance: rule/AI-assisted candidates for AutoLabel, alert-graph nodes for ExCyTIn/SecRL, and processed-trace anchors for OTRF APT29. The stored oracle prediction is used only for evaluator smoke tests.

Step units follow source-native granularity. AutoLabel uses one attacker decision or operation per step; its resulting syscalls, network fragments, library loads, and log writes remain support for that step. A new endpoint, command, target object, or intent normally opens a step; repeated attempts against the same target are merged, whereas distinct objectives are split. ExCyTIn/SecRL retains alert nodes and OTRF APT29 retains ordered event anchors. Chain-length effects are therefore interpreted within source/step-unit strata.

Reference edges also preserve source-native semantics. AutoLabel edge support is inherited from the destination step and is not an independently annotated causal proof; ExCyTIn/SecRL and OTRF edges primarily encode chronological adjacency. Accordingly, the ordering metric is temporal, not a claim of fully verified causality.

Before card generation, parsers remove reference labels and other answer-bearing source fields unavailable in the model-visible evidence, including QA answers, solution paths, incident reports, source snippets, attack scripts, and scene markers. Any ATT\&CK fields retained in the output schema are optional descriptive metadata rather than a reported benchmark label.

The input-leakage audit found no failures or warnings across 69 inputs and 12,833 cards; the artifact-consistency audit found no mismatch among any manifest row and its incident input, reference package, retrieval documents, or retrieval graph.

For each package, an author compared the derived reference with the original source record and marked every package, step, and edge as confirmed or corrected. Table~\ref{tab:app-reference-manual-check} reports the exact denominators and also covers boundaries, order, entities, support IDs, and cross-profile consistency. This was a full author audit rather than independent double annotation, so inter-annotator agreement is not reported.

\begin{table}[t]
\centering
\small
\begin{tabular}{@{}lccc@{}}
\toprule
Object & Reviewed & Confirmed & Corrected \\
\midrule
Reference packages & 69/69 & 69 & 0 \\
Reference steps & 849/849 & 849 & 0 \\
Reference edges & 780/780 & 780 & 0 \\
Boundaries/order/entities & All & All & 0 \\
Support-ID checks & All & All & 0 \\
\bottomrule
\end{tabular}
\caption{Full manual audit of the frozen reference packages against their source records.}
\label{tab:app-reference-manual-check}
\end{table}

\section{Implementation and Reproducibility Details} \label{app:agent-ecrag-config} The main paper defines the task, output schema, typed actions, agent loop, and high-level ECRAG stages. This section records only implementation details required to reproduce the frozen protocol; Figures~\ref{fig:app-prompt-base-a}--\ref{fig:app-prompt-controller-a} reproduce the complete task and controller prompts. \subsection{ECRAG Boundary and Exact Ranking} ECRAG operates only on the case's visible incident input, retrieval documents derived from its evidence cards, and the visible evidence--entity/local-adjacency index. It never reads the reference chain, oracle prediction, attack script, QA answer, source code, or any case-level gold label. \textbf{Frozen ranking rule.} Retrieval documents are indexed with scikit-learn's \texttt{TfidfVectorizer} using word unigrams and bigrams (\texttt{ngram\_range=(1,2)}) and lowercasing; all other vectorizer arguments use the scikit-learn 1.8.0 defaults recorded below.
Let $q$ be the current query and $e$ a visible evidence card. ECRAG first ranks retrieval documents by TF--IDF cosine similarity $s_{\mathrm{tfidf}}(q,e)$, retaining up to $\max(3k,12)$ positive-score candidates and taking up to $\max(k,8)$ seed identifiers. From the first eight seeds, it adds one-hop graph-neighbor evidence records and one-record same-source neighbors, each expansion capped at $2k$ identifiers. The candidate set is then reranked by
\begin{equation}
s(q,e)=s_{\mathrm{tfidf}}(q,e)+
\frac{|T_q\cap T_e|}{\sqrt{|T_q||T_e|}}+b_{\mathrm{clue}}(q,e)+b_{\mathrm{sec}}(e),
\label{eq:app-ecrag-score}
\end{equation}
where $T_q,T_e$ are normalized token sets. For each normalized card clue occurring verbatim in the normalized query,
$b_{\mathrm{clue}}$ adds $\min(0.25,0.03|T_{\mathrm{clue}}|)$. The security bonus is
\begin{equation}
b_{\mathrm{sec}}(e)=\min(3,0.25h_e+0.5v_e+2a_e+0.4r_e),
\end{equation}
where $h_e$ counts matched security-hint patterns, $v_e$ marks a high-value field, $a_e$ marks a high-signal alert, and $r_e$ marks an alert/security-incident source. Candidates are ordered by decreasing $s(q,e)$, with exact reranking ties broken by source handle, source-local line number, and numeric evidence-ID suffix. The highest $k$ cards are selected and then presented in source-local record order. Thus the chronological cue is local to each source; ECRAG does not impose a cross-source global timeline.

\subsection{Validation and Frozen Controller Configuration} The support gate blocks a submission with missing or unobserved step support while budget remains. At finalization, the grounding check removes unobserved evidence IDs, drops steps left without observed support, and removes incident edges; it never adds, relabels, or reorders surviving steps and never reads the hidden reference chain. Table~\ref{tab:app-final-agent-config} gives every frozen controller and retrieval setting used by RQ1/RQ2. The separately tagged auxiliary evidence-dependence control in Appendix~\ref{app:rq5-control} uses its recorded eight-turn protocol and is not pooled with the final panel. \begin{table}[t]
\centering
\small
\setlength{\tabcolsep}{3pt}
\begin{tabular}{lp{0.58\columnwidth}}
\toprule
Parameter & Final setting \\
\midrule
Retrieval kernel & ECRAG \\
Planner / actions & LLM planner with typed actions \\
Controller method & Reflection \\
Memory & Structured attack-chain memory with investigation state \\
Memory token budget & 12,000 \\
Memory compression & On \\
Scaffold checks & Support audit and grounding check enabled \\
Oracle mode & None \\
Prompt profile & \texttt{visible\_anchor} \\
Temperature & 0.0 \\
Sampling & One run per model--case condition \\
Turn budget & \texttt{max\_turns}=15 \\
Retrieval width & \texttt{top\_k\_per\_query}=32 \\
Card-text limit & 900 characters per returned evidence card \\
Memory cap & 48 evidence cards \\
Early stop & Stop after three no-new-evidence turns; submit the grounded chain when available \\
JSON repair & At most two repair calls per controller response \\
API retry & At most three retries for transient request failures \\
Request timeout & 900 seconds \\
\bottomrule
\end{tabular}
\caption{Final agent and retrieval configuration.}
\label{tab:app-final-agent-config}
\end{table}

\subsection{Development Range and Selection} RQ4 varied one budget at a time on R24: \(k\in\{1,4,8,16,32,48,64\}\) with \(T=15\), and \(T\in\{3,5,10,15,20,25\}\) with \(k=32\). The final \(k=32,T=15\) setting was selected as the observed multi-metric quality--cost operating point using Ret./Grp./Ord./Grd./Gap together with realized queries, tokens, and latency; Table~\ref{tab:app-rq4-budget} reports every tried row. The only paper-facing development sweeps were the \(k\) and \(T\) grids above. Each remaining controller, memory, repair, and ECRAG ranking setting was held at the single value reported in Table~\ref{tab:app-final-agent-config} and the ranking rules above, and was not selected by comparing alternative values. \subsection{Model and Inference Configuration} \begin{table*}[!t] \centering \small \setlength{\tabcolsep}{3pt} \begin{tabular}{@{}lp{0.21\textwidth}lp{0.45\textwidth}@{}} \toprule Reported name & Provider / backend ID & Reasoning & Final request settings \\ \midrule Qwen-3-32b & \texttt{ollama}/\texttt{qwen3:32b} & off & temperature 0; \texttt{num\_ctx}=40,960; \texttt{num\_predict}=16,384; JSON \\ Llama4-17b-Scout & \texttt{ollama}/\texttt{llama4:scout} & off & temperature 0; \texttt{num\_ctx}=40,960; \texttt{num\_predict}=16,384; JSON \\ DeepSeek-V4-Pro & \shortstack[l]{\texttt{deepseek}/\\[-1pt]\texttt{deepseek-v4-pro}} & off & temperature 0; \texttt{max\_tokens}=16,384; JSON object; context not sent \\ GLM-5.2 & \texttt{glm}/\texttt{glm-5.2} & off & temperature 0; \texttt{max\_tokens}=16,384; JSON object; context not sent \\ GLM-5.2-T & \texttt{glm}/\texttt{glm-5.2} & high/on & temperature 0; \texttt{max\_tokens}=16,384; JSON object; context not sent \\ GPT-5.5 & \texttt{sssaicode}/\texttt{gpt-5.5} & provider default & temperature 0; \texttt{max\_output\_tokens}=16,384 requested; JSON object; context not sent \\ \bottomrule \end{tabular} \caption{Provider/backend identifiers and model-specific request settings for the final six-configuration panel.} \label{tab:app-final-model-config} \end{table*} The two Ollama rows also retain their exact model digests and Q4\_K\_M quantization in the frozen run records. Every call used the shared system instruction \emph{You are a cybersecurity analyst. Return only the requested JSON object.} and a provider-native JSON-only response constraint. For GPT-5.5 this constraint was \texttt{text.format=json\_object}; the request also set \texttt{store=false} and \texttt{stream=true} as retention and transport controls. Others followed provider defaults. Together with Table~\ref{tab:app-final-agent-config} and the ECRAG ranking rules, this reports all researcher-controlled final inference and agent settings. \subsection{Execution Environment and Reproducibility Boundaries} DeepSeek, GLM, and GPT-5.5 ran on provider-managed hardware that was not observable. Orchestration and both Ollama models ran from the project workspace under an author account on a remote Ubuntu 22.04.2 LTS server (Linux 5.15.0-174) with dual Intel Xeon Gold 6430 CPUs, 128 logical CPUs, approximately 503.5 GiB RAM, and eight NVIDIA GeForce RTX 4090 GPUs with 24 GB memory each. The server used NVIDIA driver 580.65.06 and CUDA 13.0; the user environment used Python 3.13.5 (Miniconda), Conda 25.7.0, and Ollama 0.20.6, with NumPy 2.3.5, pandas 2.3.3, SciPy 1.16.2, scikit-learn 1.8.0, NetworkX 3.5, Matplotlib 3.10.6, and Requests 2.34.2. \subsection{Run Count, Randomness, and Frozen Artifacts} RQ1 contains one run for each of 69 cases and six configurations (414 model--case runs); RQ2 is computed from these same frozen runs and adds no model calls. RQ3 uses one run for each of 12 cases under each of three scaffold conditions (36 runs). RQ4 uses one run for each of 24 cases at 12 unique \((k,T)\) settings (288 runs); the shared \((32,15)\) baseline appears in both sweep views but was not rerun. ECRAG, schema normalization, and evaluation are deterministic for fixed artifacts and use no random seed. Model calls share no portable seed field: temperature 0 is sent to all models used. Paired case bootstraps use 10,000 resamples and frozen seed 20,260,711. Rows retain provider/model and reasoning settings, returned token counts, realized calls, latency, termination reason, artifact paths, and evaluator version. \texttt{ok\_cached} rows reuse a schema-valid result from the corresponding condition directory; separate directories and recorded row configurations prevent cross-condition reuse. The anonymous artifact includes the package-wide SHA-256 manifest, MAIN-69/R12/R24 manifests, frozen RQ1--RQ4 rows, and scripts that verify the package and rebuild paper-facing tables and figures. The separately protocol-tagged auxiliary control uses one run for each of five models, 15 cases, and the two numerical conditions C0/C1 (150 runs); Original in Figure~\ref{fig:app-rq5-conditions} is illustrative rather than a third numerical condition. This control is not used for the main claims. \subsection{Complete Task and Controller Prompts}

The final \texttt{visible\_anchor} task prompt is assembled in a fixed order: the base reconstruction contract, the runtime anti-merge block, the granularity-calibration profile, and the visible-anchor decomposition profile. Figures~\ref{fig:app-prompt-base-a}--\ref{fig:app-prompt-visible-b} reproduce these static instructions verbatim. Figure~\ref{fig:app-prompt-controller-a} gives the per-turn controller shell. Bracketed uppercase fields in that shell are not additional instructions; the runner replaces them with the case-specific visible incident stub, current turn, run-local memory, support audit, and bounded ECRAG observations. Hidden reference chains and gold-support labels are never inserted.

\section{Diagnostic Evaluator}
\label{app:metrics-funnel}

\subsection{Scorecard Metrics}

The scorecard separates evidence discovery, grouping, ordering, grounding, and final attribution. Let \(G\) be the ordered reference steps and \(P\) be the submitted steps. Each step \(s\) has a support-evidence set \(E(s)\) after unknown evidence identifiers are filtered. Let \(O\) denote the evidence observed by the agent through the retrieval loop, \(C\) denote all evidence cited in the final submitted chain, including step and causal-edge citations, and \(E_G=\bigcup_{g\in G}E(g)\) denote all reference-supporting evidence.

\textbf{Retrieval step coverage (Ret.)} measures whether retrieval exposed at least one support card for each reference step:
\[
\mathrm{Ret.}=\frac{\left|\{g\in G:E(g)\neq\emptyset \wedge E(g)\cap O\neq\emptyset\}\right|}
{\left|\{g\in G:E(g)\neq\emptyset\}\right|}.
\]

\textbf{Grouping F1 (Grp.)} is the \(B^3\) evidence-clustering F1. The evaluator projects the reference and prediction into evidence-to-step cluster labels over the same evidence universe. For each evidence item \(e\), let \(C_P(e)\) and \(C_G(e)\) be its predicted and reference clusters. The per-item \(B^3\) precision and recall are
\[
\begin{array}{ll}
P_{B^3}(e)&=\displaystyle\frac{|C_P(e)\cap C_G(e)|}{|C_P(e)|},\\
R_{B^3}(e)&=\displaystyle\frac{|C_P(e)\cap C_G(e)|}{|C_G(e)|}.
\end{array}
\]
The reported Grp. value is the harmonic mean of the mean \(B^3\) precision and mean \(B^3\) recall across evidence items. This scores merge and split errors without requiring the submitted chain to use the same number of steps as the reference chain.

\textbf{Ordering accuracy (Ord.)} and \textbf{evidence grounding F1 (Grd.)} use the same CEAF-style step alignment. Each predicted and reference step is represented as an evidence cluster. Duplicate evidence citations are assigned to the first step that cites them, unknown citations are filtered, and a one-to-one assignment maximizes total evidence overlap. We solve this assignment with the Hungarian algorithm when available, with deterministic exact or greedy fallbacks otherwise, and retain only aligned pairs with positive evidence overlap. Let \(\mathcal{A}=\{(p_k,g_k)\}_{k=1}^{m}\) be the retained aligned pairs, and let \(F_1(p,g)=2|E(p)\cap E(g)|/(|E(p)|+|E(g)|)\). Evidence grounding is
\[
\mathrm{Grd.}=
\left\{
\begin{array}{ll}
\displaystyle\frac{1}{m}\sum_{k=1}^{m}F_1(p_k,g_k), & m>0,\\[3pt]
0, & m=0.
\end{array}
\right.
\]
Because Grd. is computed only over positive-overlap aligned pairs, unmatched predicted steps are instead reflected in Grp. and the extra-step diagnostics. Grd. should therefore be interpreted jointly with Grp. and Ret. For two aligned pairs, define \(\delta^P_{ij}=\mathbf{1}[\mathrm{pos}_P(p_i)<\mathrm{pos}_P(p_j)]\) and \(\delta^G_{ij}=\mathbf{1}[\mathrm{pos}_G(g_i)<\mathrm{pos}_G(g_j)]\). Ordering accuracy is
\[
\mathrm{Ord.}=
\left\{
\begin{array}{ll}
\displaystyle\frac{2}{m(m-1)}
\sum_{1\le i<j\le m}\mathbf{1}[\delta^P_{ij}=\delta^G_{ij}], & m\ge 2,\\[3pt]
1, & m=1,\\
0, & m=0.
\end{array}
\right.
\]
In RQ2, a reference step is assigned to E4 if it participates in at least one incorrectly ordered aligned-step pair. Because E4 is a step-level failure label whereas Ord. is a pair-level case score, the E4 rate need not equal \(1-\mathrm{Ord}\).

\textbf{Attribution gap rate (Gap)} measures how often the agent observed reference-supporting evidence but omitted it from the final submitted chain:
\[
\mathrm{Gap}=\frac{|(O\cap E_G)\setminus C|}{|O\cap E_G|}.
\]
A model can achieve high Ret. yet still have a high Gap when it retrieves support for many reference steps but omits much of that observed support from the final chain.

     \subsection{Aggregation} Main metrics are case-macro means, giving each case equal weight; RQ2 reports proportions over coverable reference steps. Subgroup summaries use the source, evidence-profile, or length strata stated in each table, and RQ3/RQ4 are reported descriptively with their exact sample sizes. Schema validity, step counts, citation errors, merge/split diagnostics, and resource use are retained only for auditing. \subsection{Progressive Failure Funnel (RQ2)} For each coverable reference step \(g\), let \(O_g=E(g)\cap O\) and \(U_g=O_g\cap C_s\), where \(C_s\) contains evidence cited by final submitted steps. Table~\ref{tab:app-funnel-gates} applies the main paper's E1--E4/OK stages in order. Causal-edge-only citations do not count as step attribution; Gap, which uses both step and edge citations, is reported separately. \begin{table*}[!t]
\centering
\small
\setlength{\tabcolsep}{4pt}
\begin{tabular}{lp{0.34\textwidth}p{0.49\textwidth}}
\toprule
Label & Formal gate & Interpretation \\
\midrule
\colorbox{appBlueFill}{\strut\textbf{E1}} & \(O_g=\emptyset\) & No visible supporting evidence for the reference step was observed by the agent \\
\colorbox{appAmberFill}{\strut\textbf{E2}} & \(O_g\neq\emptyset\land U_g=\emptyset\) & Supporting evidence was observed but no submitted step cited it \\
\colorbox{appRoseFill}{\strut\textbf{E3}} & \(\emptyset\neq U_g\subset O_g\) & A submitted step cited only part of the observed support \\
\colorbox{appVioletFill}{\strut\textbf{E4}} & \(U_g=O_g\) and \(g\) participates in at least one incorrectly ordered aligned-step pair & All observed support was cited, but the step participates in at least one incorrectly ordered aligned-step pair \\
\colorbox{appGreenFill}{\strut\textbf{OK}} & none of E1--E4 holds & The step passes E1--E4: its observed support was fully cited, and no aligned-order violation was assigned to it \\
\bottomrule
\end{tabular}
\caption{Progressive RQ2 decision order over coverable reference steps, defined as steps with at least one supporting item in the model-visible evidence set. A step is assigned once, at the first true condition from top to bottom.}
\label{tab:app-funnel-gates}
\end{table*}

\subsection{Manual Validation of Evaluator Outputs}
We constructed a blinded, purposive diagnostic sample from 12 frozen MAIN-69 case-level evaluator outputs. The sampled artifacts use the same frozen scorecard and evidence-alignment implementation as the reported runs. This was not a probability sample: cases were selected to span the three source families, clean/noisy/raw evidence profiles, S/M/L length buckets, and eight automatic success/failure patterns; Table~\ref{tab:app-evaluator-manual-check} gives the exact distribution.

Reviewer-facing cards hid model and case identities but retained the automatic metrics and failure labels because agreement with those outputs was the object of review. For each case, reviewers compared the reference and predicted chains and evidence excerpts, then recorded five judgments in the applicable Figure~1 order: unsupported-output handling, step alignment/grouping, ordering, evidence grounding, and the dominant failure label. All five judgments agreed for every case (60/60), and no output or label was corrected. This is a diagnostic validation of the sampled outputs, not a full-corpus agreement estimate.

\begin{table}[t]
\centering
\small
\setlength{\tabcolsep}{4pt}
\begin{tabular}{@{}p{0.25\columnwidth}p{0.68\columnwidth}@{}}
\toprule
Sample factor & Distribution ($n$) \\
\midrule
Source family & AutoLabel 7; ExCyTIn-Bench 4; OTRF APT29 1 \\
Evidence profile & Clean 5; noisy 3; raw 4 \\
Length bucket & S 7; M 2; L 3 \\
Diagnostic pattern & Attribution gap 3; retrieval miss 2; merge 1; split 1; grounding 1; ordering 1; low-quality 1; success/minor 2 \\
Sampling scheme & Purposive diagnostic sample; non-random \\
\bottomrule
\end{tabular}

\vspace{4pt}
\begin{tabular}{@{}p{0.54\columnwidth}cc@{}}
\toprule
Review judgment & Confirmed & Corrected \\
\midrule
Unsupported-output handling & 12/12 & 0 \\
Step alignment/grouping & 12/12 & 0 \\
Ordering & 12/12 & 0 \\
Evidence grounding & 12/12 & 0 \\
Dominant failure label & 12/12 & 0 \\
\midrule
All five review fields & 60/60 (100\%) & 0 \\
\bottomrule
\end{tabular}
\caption{Sampling coverage and manual agreement for the 12-case evaluator audit.}
\label{tab:app-evaluator-manual-check}
\end{table}

\section{Qualitative Cases and Failure Traces}
\label{app:qualitative-cases}

This section follows one frozen RQ1 run through retrieval, submission, alignment, and the final RQ2 gate. We choose a near-miss rather than a collapse: the predicted and reference chains have the same length and identical one-to-one evidence coverage, but a local permutation still produces an ordering failure. The trace explains the mechanism and does not re-estimate its frequency.

\subsection{End-to-End Near-Miss: GPT-5.5 on Inc.-134 Clean}

The reference and submitted chains both contain nine steps. Turn 1 retrieves all nine gold-support cards; turn 2 submits nine steps, cites every observed card exactly once, and leaves no attribution gap. Ret., Grp., and Grd. are therefore 100.0 and Gap is 0.0. Only Ord. is imperfect at 94.4.

\noindent\begin{minipage}{\columnwidth}
\appcolpanel{appBlueRule}{appBlueFill}{%
\textbf{Frozen reference order.}\par
\texttt{E0001} $\rightarrow$ \texttt{E0002} $\rightarrow$ \texttt{E0003} $\rightarrow$ \texttt{E0004} $\rightarrow$ \texttt{E0005}\par
\texttt{E0006} $\rightarrow$ \texttt{E0007} $\rightarrow$ \texttt{E0008} $\rightarrow$ \texttt{E0009}}
\vspace{2pt}
\appcolpanel{appRoseRule}{appRoseFill}{%
\textbf{Submitted order.}\par
\texttt{E0001} $\rightarrow$ \texttt{E0002} $\rightarrow$ \textcolor{appRoseRule}{\texttt{E0005}} $\rightarrow$ \texttt{E0003} $\rightarrow$ \texttt{E0004}\par
\texttt{E0006} $\rightarrow$ \texttt{E0007} $\rightarrow$ \texttt{E0008} $\rightarrow$ \texttt{E0009}\par
The password-spray step supported by \texttt{E0005} moves from reference position 5 to submitted position 3; every other evidence anchor keeps its relative order.}
\end{minipage}

\begin{figure*}[!t]
\centering
\includegraphics[width=0.72\textwidth]{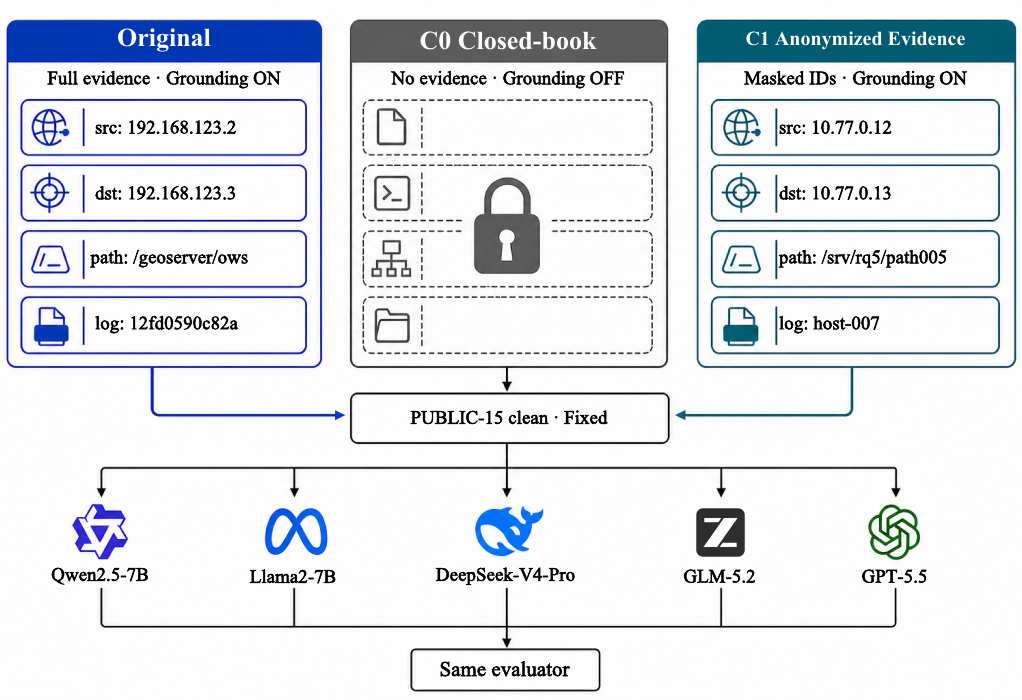}
\caption{Auxiliary evidence conditions. Original provides the full evidence environment, C0 removes evidence and disables grounding, and C1 anonymizes surface identifiers while preserving evidence structure and grounding. The same evaluator is used across conditions.}
\label{fig:app-rq5-conditions}
\end{figure*}

\noindent\begin{minipage}{\columnwidth}
\textbf{Complete reference-to-submission alignment.}\par\smallskip
\centering
\scriptsize
\setlength{\tabcolsep}{1.7pt}
\renewcommand{\arraystretch}{0.94}
\begin{tabular}{@{}lp{0.28\columnwidth}lp{0.28\columnwidth}l@{}}
\toprule
Ref. & Reference action & Pred. & Submitted action & Gate \\
\midrule
G001/\texttt{E0001} & Malicious URL email & S1 & Deliver malicious URL email & OK \\
G002/\texttt{E0002} & Malicious URL click & S2 & Click malicious URL & OK \\
G003/\texttt{E0003} & Anonymous-IP sign-in & S4 & Sign in from anonymous IP & \textbf{E4} \\
G004/\texttt{E0004} & Malicious-IP sign-in & S5 & Sign in from malicious IP & \textbf{E4} \\
G005/\texttt{E0005} & Password spray & S3 & Conduct password spray & \textbf{E4} \\
G006/\texttt{E0006} & Account compromised & S6 & Compromise after spray & OK \\
G007/\texttt{E0007} & Inbox-rule manipulation & S7 & Manipulate inbox rule & OK \\
G008/\texttt{E0008} & BEC financial fraud & S8 & Create email-hiding rule & OK \\
G009/\texttt{E0009} & Suspicious BEC email & S9 & Send BEC-related emails & OK \\
\bottomrule
\end{tabular}
\par\smallskip
\footnotesize All nine evidence clusters align one-to-one; only G003--G005 participate in the local permutation.
\end{minipage}

The full-chain alignment traces every step rather than showing only the offending pair. G001--G002 and G006--G009 retain both content and order. G003--G005 are all assigned E4 because their aligned prediction positions are 4, 5, and 3. Moving G005 ahead of G003--G004 creates exactly two inverted reference pairs out of 36, giving $34/36=94.4\%$ pairwise order accuracy. The aligned chain has edit distance 2 and an eight-step longest common subsequence.

The controller's support audit does not block submission: all nine cards are observed and cited, every submitted step is supported, and all eight submitted edges have endpoint evidence. This is precisely why the error reaches E4 rather than an earlier gate. The example adds information absent from the aggregate scorecard: a run can recover the complete event set, preserve one-to-one grouping and grounding, and still fail chain reconstruction through a single local placement decision.

For boundary comparison, the Qwen-3-32b APT29 trace fails earlier because G008 support \texttt{E0128} is never observed (E1), while the Llama4-17b-Scout Inc.-39 trace observes \texttt{E0124} and \texttt{E0126} but never cites them (E2). At the other end, GPT-5.5 on Inc.-55 clean preserves all 27 reference steps in order and receives OK throughout. These shorter contrasts locate the expanded near-miss within the complete funnel without duplicating another full audit.

\section{Additional Experimental Results}
\label{app:additional-results}

This section gives the complete RQ1 case grid, the exact values behind the RQ2 diagnostic plots, the complete RQ4 budget sweeps, and a separately tagged auxiliary evidence-dependence control. These tables complement, rather than replace, the MAIN-69 case-level results in the main paper.

\begingroup
\let\section\subsection
\section{Complete RQ1 Case-Level Results}
\label{app:rq1-complete-case-results}

The following compact vertical tables report all MAIN-69 RQ1 case-model results. Each panel column block contains several benchmark cases; model names are rotated to fit all six systems. Case/GT gives the compact case ID and reference-chain length; Pred. is the submitted-chain length. Ret., Grp., Ord., Grd., and Gap denote the five diagnostic metrics used in the main text.

\begin{table*}[t]
\centering
\scriptsize
\setlength{\tabcolsep}{1.25pt}
\begin{tabular}{@{}llcccccc@{\hspace{2pt}}llcccccc@{\hspace{2pt}}llcccccc@{}}
\toprule
Case/GT & Metric & \rotatebox{90}{Qwen-3-32b} & \rotatebox{90}{Llama4-17b-Scout} & \rotatebox{90}{DeepSeek-V4-Pro} & \rotatebox{90}{GLM-5.2} & \rotatebox{90}{GLM-5.2-T} & \rotatebox{90}{GPT-5.5} & Case/GT & Metric & \rotatebox{90}{Qwen-3-32b} & \rotatebox{90}{Llama4-17b-Scout} & \rotatebox{90}{DeepSeek-V4-Pro} & \rotatebox{90}{GLM-5.2} & \rotatebox{90}{GLM-5.2-T} & \rotatebox{90}{GPT-5.5} & Case/GT & Metric & \rotatebox{90}{Qwen-3-32b} & \rotatebox{90}{Llama4-17b-Scout} & \rotatebox{90}{DeepSeek-V4-Pro} & \rotatebox{90}{GLM-5.2} & \rotatebox{90}{GLM-5.2-T} & \rotatebox{90}{GPT-5.5} \\
\cmidrule(lr){1-8}\cmidrule(lr){9-16}\cmidrule(lr){17-24}
S01/3 & Pred. & 4 & 2 & 3 & 3 & 3 & 3 & S07/2 & Pred. & 2 & 2 & 2 & 2 & 2 & 1 & S13/2 & Pred. & 8 & 2 & 8 & 8 & 3 & 8 \\
 & $\Delta$ & +1 & -1 & 0 & 0 & 0 & 0 &  & $\Delta$ & 0 & 0 & 0 & 0 & 0 & -1 &  & $\Delta$ & +6 & 0 & +6 & +6 & +1 & +6 \\
 & Ret. & 1.00 & 1.00 & 1.00 & 1.00 & 1.00 & 1.00 &  & Ret. & 1.00 & 1.00 & 1.00 & 1.00 & 1.00 & 1.00 &  & Ret. & 1.00 & 1.00 & 1.00 & 1.00 & 1.00 & 1.00 \\
 & Grp. & 0.86 & 0.86 & 1.00 & 0.68 & 0.86 & 0.68 &  & Grp. & 1.00 & 1.00 & 1.00 & 1.00 & 1.00 & 0.67 &  & Grp. & 0.35 & 0.56 & 0.35 & 0.35 & 0.67 & 0.35 \\
 & Ord. & 1.00 & 1.00 & 1.00 & 1.00 & 1.00 & 1.00 &  & Ord. & 1.00 & 1.00 & 1.00 & 1.00 & 1.00 & 1.00 &  & Ord. & 1.00 & 1.00 & 1.00 & 1.00 & 1.00 & 1.00 \\
 & Grd. & 0.93 & 0.83 & 1.00 & 0.63 & 0.89 & 0.63 &  & Grd. & 1.00 & 1.00 & 1.00 & 1.00 & 1.00 & 0.67 &  & Grd. & 0.29 & 0.67 & 0.29 & 0.29 & 0.80 & 0.29 \\
 & Gap & 0.08 & 0.00 & 0.00 & 0.58 & 0.25 & 0.58 &  & Gap & 0.00 & 0.00 & 0.00 & 0.00 & 0.00 & 0.00 &  & Gap & 0.33 & 0.50 & 0.00 & 0.00 & 0.00 & 0.33 \\
\addlinespace[0.4pt]
S02/3 & Pred. & 4 & 2 & 5 & 6 & 6 & 6 & S08/2 & Pred. & 6 & 2 & 3 & 10 & 9 & 10 & S14/2 & Pred. & 6 & 2 & 48 & 20 & 5 & 9 \\
 & $\Delta$ & +1 & -1 & +2 & +3 & +3 & +3 &  & $\Delta$ & +4 & 0 & +1 & +8 & +7 & +8 &  & $\Delta$ & +4 & 0 & +46 & +18 & +3 & +7 \\
 & Ret. & 0.33 & 1.00 & 1.00 & 0.67 & 0.67 & 1.00 &  & Ret. & 0.50 & 1.00 & 1.00 & 1.00 & 1.00 & 1.00 &  & Ret. & 1.00 & 1.00 & 1.00 & 1.00 & 1.00 & 1.00 \\
 & Grp. & 0.79 & 0.65 & 0.61 & 0.86 & 0.75 & 0.65 &  & Grp. & 0.53 & 0.67 & 0.86 & 0.45 & 0.50 & 0.45 &  & Grp. & 0.63 & 0.56 & 0.12 & 0.29 & 0.53 & 0.55 \\
 & Ord. & 1.00 & 1.00 & 1.00 & 1.00 & 1.00 & 1.00 &  & Ord. & 0.00 & 1.00 & 1.00 & 1.00 & 1.00 & 1.00 &  & Ord. & 1.00 & 1.00 & 1.00 & 1.00 & 1.00 & 1.00 \\
 & Grd. & 0.92 & 0.17 & 0.78 & 1.00 & 1.00 & 0.63 &  & Grd. & 0.00 & 0.39 & 1.00 & 1.00 & 1.00 & 1.00 &  & Grd. & 0.29 & 0.48 & 0.29 & 0.29 & 0.53 & 0.80 \\
 & Gap & 0.00 & 0.88 & 0.20 & 0.00 & 0.00 & 0.58 &  & Gap & 1.00 & 0.00 & 0.50 & 0.00 & 0.00 & 0.00 &  & Gap & 0.75 & 0.50 & 0.00 & 0.33 & 0.00 & 0.33 \\
\addlinespace[0.4pt]
S03/3 & Pred. & 6 & 2 & 3 & 6 & 5 & 8 & S09/2 & Pred. & 6 & 2 & 12 & 7 & 5 & 21 & S15/2 & Pred. & 18 & 2 & 17 & 12 & 4 & 2 \\
 & $\Delta$ & +3 & -1 & 0 & +3 & +2 & +5 &  & $\Delta$ & +4 & 0 & +10 & +5 & +3 & +19 &  & $\Delta$ & +16 & 0 & +15 & +10 & +2 & 0 \\
 & Ret. & 0.33 & 0.33 & 1.00 & 1.00 & 1.00 & 1.00 &  & Ret. & 0.00 & 0.00 & 1.00 & 1.00 & 1.00 & 1.00 &  & Ret. & 1.00 & 1.00 & 1.00 & 1.00 & 1.00 & 1.00 \\
 & Grp. & 0.67 & 0.72 & 0.67 & 0.90 & 0.64 & 0.61 &  & Grp. & 0.50 & 0.73 & 0.38 & 0.52 & 0.61 & 0.19 &  & Grp. & 0.28 & 0.69 & 0.30 & 0.36 & 0.57 & 0.53 \\
 & Ord. & 1.00 & 0.00 & 1.00 & 1.00 & 1.00 & 1.00 &  & Ord. & 0.00 & 0.00 & 1.00 & 1.00 & 1.00 & 1.00 &  & Ord. & 1.00 & 1.00 & 1.00 & 1.00 & 1.00 & 1.00 \\
 & Grd. & 0.77 & 0.00 & 0.62 & 0.75 & 0.61 & 0.63 &  & Grd. & 0.00 & 0.00 & 1.00 & 1.00 & 1.00 & 1.00 &  & Grd. & 0.29 & 0.67 & 0.34 & 0.29 & 0.71 & 0.56 \\
 & Gap & 0.00 & 1.00 & 0.58 & 0.00 & 0.29 & 0.58 &  & Gap & 0.00 & 0.00 & 0.00 & 0.00 & 0.00 & 0.00 &  & Gap & 0.33 & 0.67 & 0.00 & 0.00 & 0.25 & 0.33 \\
\addlinespace[0.4pt]
S04/2 & Pred. & 1 & 1 & 1 & 1 & 1 & 4 & S10/3 & Pred. & 10 & 3 & 4 & 7 & 3 & 3 & S16/2 & Pred. & 1 & 2 & 1 & 1 & 1 & 1 \\
 & $\Delta$ & -1 & -1 & -1 & -1 & -1 & +2 &  & $\Delta$ & +7 & 0 & +1 & +4 & 0 & 0 &  & $\Delta$ & -1 & 0 & -1 & -1 & -1 & -1 \\
 & Ret. & 1.00 & 1.00 & 1.00 & 1.00 & 1.00 & 1.00 &  & Ret. & 1.00 & 1.00 & 1.00 & 1.00 & 1.00 & 1.00 &  & Ret. & 1.00 & 1.00 & 1.00 & 1.00 & 1.00 & 1.00 \\
 & Grp. & 0.71 & 0.71 & 0.71 & 1.00 & 1.00 & 0.71 &  & Grp. & 0.51 & 0.71 & 0.96 & 0.62 & 0.86 & 0.78 &  & Grp. & 0.67 & 0.67 & 1.00 & 1.00 & 1.00 & 1.00 \\
 & Ord. & 1.00 & 1.00 & 1.00 & 1.00 & 1.00 & 1.00 &  & Ord. & 1.00 & 0.67 & 1.00 & 1.00 & 1.00 & 1.00 &  & Ord. & 1.00 & 1.00 & 1.00 & 1.00 & 1.00 & 1.00 \\
 & Grd. & 0.50 & 0.80 & 0.50 & 1.00 & 1.00 & 0.80 &  & Grd. & 0.15 & 0.61 & 0.93 & 0.77 & 0.80 & 0.58 &  & Grd. & 0.67 & 0.67 & 1.00 & 1.00 & 1.00 & 1.00 \\
 & Gap & 0.67 & 0.33 & 0.67 & 0.00 & 0.00 & 0.33 &  & Gap & 0.44 & 0.39 & 0.00 & 0.00 & 0.50 & 0.72 &  & Gap & 0.50 & 0.00 & 0.00 & 0.00 & 0.00 & 0.00 \\
\addlinespace[0.4pt]
S05/2 & Pred. & 20 & 1 & 2 & 2 & 2 & 4 & S11/3 & Pred. & 10 & 3 & 5 & 11 & 6 & 5 & S17/2 & Pred. & 7 & 1 & 8 & 6 & 4 & 8 \\
 & $\Delta$ & +18 & -1 & 0 & 0 & 0 & +2 &  & $\Delta$ & +7 & 0 & +2 & +8 & +3 & +2 &  & $\Delta$ & +5 & -1 & +6 & +4 & +2 & +6 \\
 & Ret. & 1.00 & 1.00 & 1.00 & 1.00 & 1.00 & 1.00 &  & Ret. & 1.00 & 1.00 & 1.00 & 1.00 & 1.00 & 1.00 &  & Ret. & 1.00 & 1.00 & 1.00 & 1.00 & 1.00 & 1.00 \\
 & Grp. & 0.18 & 1.00 & 0.67 & 1.00 & 1.00 & 0.85 &  & Grp. & 0.60 & 0.70 & 0.64 & 0.62 & 0.71 & 0.92 &  & Grp. & 0.40 & 0.81 & 0.50 & 0.48 & 0.70 & 0.50 \\
 & Ord. & 1.00 & 0.00 & 1.00 & 1.00 & 1.00 & 1.00 &  & Ord. & 1.00 & 0.67 & 1.00 & 1.00 & 1.00 & 1.00 &  & Ord. & 1.00 & 1.00 & 1.00 & 1.00 & 1.00 & 1.00 \\
 & Grd. & 0.50 & 0.00 & 0.40 & 1.00 & 1.00 & 0.80 &  & Grd. & 0.58 & 0.49 & 0.83 & 0.63 & 0.63 & 0.72 &  & Grd. & 0.67 & 0.25 & 1.00 & 0.67 & 0.67 & 1.00 \\
 & Gap & 0.00 & 1.00 & 0.67 & 0.00 & 0.00 & 0.00 &  & Gap & 0.15 & 0.50 & 0.24 & 0.00 & 0.60 & 0.61 &  & Gap & 0.50 & 0.50 & 0.00 & 0.00 & 0.00 & 0.00 \\
\addlinespace[0.4pt]
S06/2 & Pred. & 1 & 1 & 2 & 4 & 1 & 2 & S12/3 & Pred. & 4 & 3 & 5 & 10 & 9 & 6 & S18/2 & Pred. & 6 & 1 & 1 & 4 & 5 & 9 \\
 & $\Delta$ & -1 & -1 & 0 & +2 & -1 & 0 &  & $\Delta$ & +1 & 0 & +2 & +7 & +6 & +3 &  & $\Delta$ & +4 & -1 & -1 & +2 & +3 & +7 \\
 & Ret. & 0.00 & 0.00 & 0.00 & 0.00 & 0.00 & 0.00 &  & Ret. & 0.33 & 1.00 & 1.00 & 1.00 & 1.00 & 1.00 &  & Ret. & 1.00 & 1.00 & 0.00 & 0.00 & 0.00 & 1.00 \\
 & Grp. & 1.00 & 1.00 & 0.92 & 0.70 & 1.00 & 0.75 &  & Grp. & 0.70 & 0.84 & 0.77 & 0.56 & 0.65 & 0.71 &  & Grp. & 0.44 & 1.00 & 1.00 & 0.54 & 0.38 & 0.35 \\
 & Ord. & 0.00 & 0.00 & 0.00 & 0.00 & 0.00 & 0.00 &  & Ord. & 0.00 & 1.00 & 1.00 & 1.00 & 1.00 & 1.00 &  & Ord. & 1.00 & 0.00 & 0.00 & 0.00 & 0.00 & 1.00 \\
 & Grd. & 0.00 & 0.00 & 0.00 & 0.00 & 0.00 & 0.00 &  & Grd. & 0.00 & 1.00 & 0.58 & 0.52 & 0.85 & 0.62 &  & Grd. & 0.67 & 0.00 & 0.00 & 0.00 & 0.00 & 0.67 \\
 & Gap & 0.00 & 0.00 & 0.00 & 0.00 & 0.00 & 0.00 &  & Gap & 1.00 & 0.67 & 0.69 & 0.56 & 0.00 & 0.61 &  & Gap & 0.50 & 1.00 & 0.00 & 0.00 & 0.00 & 0.50 \\
\bottomrule
\end{tabular}
\caption{Complete RQ1 case-level results (1/4). This vertical panel covers S01--S18. Case/GT gives the compact case ID and reference-chain length; $\Delta=\mathrm{Pred.}-\mathrm{GT}$. Metric abbreviations follow the main text.}
\label{tab:app-rq1-complete-1}
\end{table*}

\begin{table*}[t]
\centering
\scriptsize
\setlength{\tabcolsep}{1.25pt}
\begin{tabular}{@{}llcccccc@{\hspace{2pt}}llcccccc@{\hspace{2pt}}llcccccc@{}}
\toprule
Case/GT & Metric & \rotatebox{90}{Qwen-3-32b} & \rotatebox{90}{Llama4-17b-Scout} & \rotatebox{90}{DeepSeek-V4-Pro} & \rotatebox{90}{GLM-5.2} & \rotatebox{90}{GLM-5.2-T} & \rotatebox{90}{GPT-5.5} & Case/GT & Metric & \rotatebox{90}{Qwen-3-32b} & \rotatebox{90}{Llama4-17b-Scout} & \rotatebox{90}{DeepSeek-V4-Pro} & \rotatebox{90}{GLM-5.2} & \rotatebox{90}{GLM-5.2-T} & \rotatebox{90}{GPT-5.5} & Case/GT & Metric & \rotatebox{90}{Qwen-3-32b} & \rotatebox{90}{Llama4-17b-Scout} & \rotatebox{90}{DeepSeek-V4-Pro} & \rotatebox{90}{GLM-5.2} & \rotatebox{90}{GLM-5.2-T} & \rotatebox{90}{GPT-5.5} \\
\cmidrule(lr){1-8}\cmidrule(lr){9-16}\cmidrule(lr){17-24}
S19/3 & Pred. & 5 & 2 & 6 & 5 & 3 & 4 & S25/2 & Pred. & 10 & 2 & 2 & 7 & 2 & 3 & S31/2 & Pred. & 14 & 2 & 3 & 8 & 4 & 2 \\
 & $\Delta$ & +2 & -1 & +3 & +2 & 0 & +1 &  & $\Delta$ & +8 & 0 & 0 & +5 & 0 & +1 &  & $\Delta$ & +12 & 0 & +1 & +6 & +2 & 0 \\
 & Ret. & 1.00 & 1.00 & 1.00 & 1.00 & 1.00 & 1.00 &  & Ret. & 1.00 & 1.00 & 1.00 & 1.00 & 1.00 & 1.00 &  & Ret. & 1.00 & 1.00 & 1.00 & 1.00 & 1.00 & 1.00 \\
 & Grp. & 0.58 & 0.63 & 0.73 & 0.65 & 0.90 & 0.89 &  & Grp. & 0.56 & 0.71 & 0.62 & 0.44 & 0.62 & 0.65 &  & Grp. & 0.38 & 0.71 & 0.68 & 0.45 & 0.83 & 0.65 \\
 & Ord. & 1.00 & 1.00 & 1.00 & 1.00 & 1.00 & 1.00 &  & Ord. & 1.00 & 1.00 & 1.00 & 1.00 & 1.00 & 1.00 &  & Ord. & 0.00 & 1.00 & 1.00 & 1.00 & 1.00 & 1.00 \\
 & Grd. & 0.73 & 0.51 & 0.86 & 0.75 & 0.76 & 0.72 &  & Grd. & 0.15 & 0.67 & 0.40 & 0.39 & 0.40 & 0.33 &  & Grd. & 0.15 & 0.67 & 0.41 & 0.54 & 0.83 & 0.28 \\
 & Gap & 0.11 & 0.44 & 0.00 & 0.00 & 0.56 & 0.56 &  & Gap & 0.67 & 0.75 & 0.75 & 0.42 & 0.75 & 0.75 &  & Gap & 0.42 & 0.75 & 0.71 & 0.08 & 0.00 & 0.83 \\
\addlinespace[0.4pt]
S20/3 & Pred. & 4 & 3 & 3 & 5 & 5 & 4 & S26/2 & Pred. & 3 & 3 & 2 & 7 & 5 & 5 & S32/2 & Pred. & 3 & 3 & 3 & 5 & 3 & 5 \\
 & $\Delta$ & +1 & 0 & 0 & +2 & +2 & +1 &  & $\Delta$ & +1 & +1 & 0 & +5 & +3 & +3 &  & $\Delta$ & +1 & +1 & +1 & +3 & +1 & +3 \\
 & Ret. & 1.00 & 1.00 & 1.00 & 1.00 & 1.00 & 1.00 &  & Ret. & 1.00 & 0.50 & 1.00 & 1.00 & 1.00 & 0.50 &  & Ret. & 1.00 & 0.50 & 1.00 & 0.50 & 1.00 & 1.00 \\
 & Grp. & 0.80 & 0.68 & 0.71 & 0.74 & 0.63 & 0.91 &  & Grp. & 0.76 & 0.74 & 0.74 & 0.60 & 0.69 & 0.66 &  & Grp. & 0.75 & 0.67 & 0.73 & 0.68 & 0.62 & 0.74 \\
 & Ord. & 1.00 & 1.00 & 1.00 & 1.00 & 1.00 & 1.00 &  & Ord. & 0.00 & 0.00 & 1.00 & 1.00 & 1.00 & 1.00 &  & Ord. & 1.00 & 1.00 & 1.00 & 1.00 & 1.00 & 1.00 \\
 & Grd. & 0.50 & 0.22 & 0.52 & 0.72 & 0.70 & 0.76 &  & Grd. & 0.00 & 0.00 & 0.36 & 0.19 & 0.40 & 0.27 &  & Grd. & 0.06 & 0.22 & 0.15 & 0.40 & 0.37 & 0.15 \\
 & Gap & 0.67 & 0.50 & 0.61 & 0.44 & 0.44 & 0.56 &  & Gap & 1.00 & 1.00 & 0.00 & 0.40 & 0.88 & 0.33 &  & Gap & 0.89 & 0.60 & 0.90 & 0.40 & 0.54 & 0.89 \\
\addlinespace[0.4pt]
S21/3 & Pred. & 5 & 2 & 1 & 3 & 5 & 3 & S27/2 & Pred. & 3 & 2 & 8 & 2 & 8 & 2 & S33/2 & Pred. & 5 & 2 & 4 & 8 & 10 & 5 \\
 & $\Delta$ & +2 & -1 & -2 & 0 & +2 & 0 &  & $\Delta$ & +1 & 0 & +6 & 0 & +6 & 0 &  & $\Delta$ & +3 & 0 & +2 & +6 & +8 & +3 \\
 & Ret. & 0.00 & 0.00 & 0.00 & 0.67 & 0.67 & 1.00 &  & Ret. & 0.50 & 0.50 & 1.00 & 1.00 & 1.00 & 1.00 &  & Ret. & 0.50 & 0.50 & 1.00 & 1.00 & 0.50 & 1.00 \\
 & Grp. & 0.70 & 0.75 & 0.71 & 0.79 & 0.80 & 0.87 &  & Grp. & 0.88 & 0.75 & 0.63 & 0.74 & 0.53 & 0.71 &  & Grp. & 0.71 & 0.71 & 0.65 & 0.58 & 0.51 & 0.62 \\
 & Ord. & 0.00 & 0.00 & 0.00 & 1.00 & 1.00 & 1.00 &  & Ord. & 0.00 & 0.00 & 1.00 & 1.00 & 1.00 & 1.00 &  & Ord. & 0.00 & 1.00 & 1.00 & 1.00 & 1.00 & 1.00 \\
 & Grd. & 0.00 & 0.00 & 0.00 & 0.65 & 0.80 & 0.71 &  & Grd. & 0.00 & 0.00 & 0.21 & 0.64 & 0.46 & 0.40 &  & Grd. & 0.00 & 0.67 & 0.49 & 0.39 & 0.59 & 0.34 \\
 & Gap & 0.00 & 0.00 & 0.00 & 0.00 & 0.00 & 0.56 &  & Gap & 1.00 & 1.00 & 0.88 & 0.41 & 0.67 & 0.88 &  & Gap & 1.00 & 0.00 & 0.11 & 0.73 & 0.17 & 0.67 \\
\addlinespace[0.4pt]
S22/2 & Pred. & 12 & 2 & 5 & 7 & 3 & 2 & S28/2 & Pred. & 4 & 2 & 7 & 8 & 5 & 2 & S34/4 & Pred. & 5 & 3 & 6 & 9 & 11 & 15 \\
 & $\Delta$ & +10 & 0 & +3 & +5 & +1 & 0 &  & $\Delta$ & +2 & 0 & +5 & +6 & +3 & 0 &  & $\Delta$ & +1 & -1 & +2 & +5 & +7 & +11 \\
 & Ret. & 1.00 & 1.00 & 1.00 & 1.00 & 1.00 & 1.00 &  & Ret. & 1.00 & 1.00 & 1.00 & 1.00 & 1.00 & 1.00 &  & Ret. & 1.00 & 1.00 & 0.75 & 0.50 & 1.00 & 1.00 \\
 & Grp. & 0.70 & 0.71 & 0.79 & 0.52 & 0.57 & 0.64 &  & Grp. & 0.66 & 0.73 & 0.53 & 0.53 & 0.56 & 0.93 &  & Grp. & 0.62 & 0.56 & 0.51 & 0.44 & 0.34 & 0.24 \\
 & Ord. & 1.00 & 1.00 & 1.00 & 1.00 & 1.00 & 1.00 &  & Ord. & 1.00 & 1.00 & 1.00 & 1.00 & 1.00 & 1.00 &  & Ord. & 1.00 & 1.00 & 0.00 & 1.00 & 1.00 & 1.00 \\
 & Grd. & 0.15 & 0.67 & 0.75 & 0.58 & 0.45 & 0.34 &  & Grd. & 0.64 & 0.83 & 0.75 & 0.75 & 0.75 & 0.58 &  & Grd. & 0.33 & 0.45 & 0.00 & 0.67 & 0.72 & 0.56 \\
 & Gap & 0.08 & 0.75 & 0.21 & 0.00 & 0.62 & 0.79 &  & Gap & 0.43 & 0.43 & 0.00 & 0.00 & 0.00 & 0.79 &  & Gap & 0.25 & 0.00 & 1.00 & 0.00 & 0.00 & 0.00 \\
\addlinespace[0.4pt]
S23/2 & Pred. & 2 & 2 & 5 & 8 & 8 & 2 & S29/2 & Pred. & 2 & 3 & 10 & 13 & 8 & 9 & S35/4 & Pred. & 6 & 3 & 4 & 6 & 4 & 5 \\
 & $\Delta$ & 0 & 0 & +3 & +6 & +6 & 0 &  & $\Delta$ & 0 & +1 & +8 & +11 & +6 & +7 &  & $\Delta$ & +2 & -1 & 0 & +2 & 0 & +1 \\
 & Ret. & 1.00 & 0.50 & 1.00 & 1.00 & 0.50 & 0.50 &  & Ret. & 1.00 & 1.00 & 1.00 & 1.00 & 1.00 & 1.00 &  & Ret. & 1.00 & 1.00 & 1.00 & 1.00 & 1.00 & 1.00 \\
 & Grp. & 0.68 & 0.75 & 0.67 & 0.56 & 0.62 & 0.70 &  & Grp. & 0.92 & 0.80 & 0.68 & 0.38 & 0.58 & 0.72 &  & Grp. & 1.00 & 0.86 & 1.00 & 1.00 & 1.00 & 1.00 \\
 & Ord. & 0.00 & 0.00 & 1.00 & 1.00 & 1.00 & 1.00 &  & Ord. & 1.00 & 1.00 & 1.00 & 1.00 & 1.00 & 1.00 &  & Ord. & 0.83 & 1.00 & 1.00 & 1.00 & 1.00 & 1.00 \\
 & Grd. & 0.00 & 0.00 & 0.63 & 0.38 & 0.74 & 0.29 &  & Grd. & 0.80 & 0.25 & 0.50 & 0.75 & 0.62 & 0.58 &  & Grd. & 1.00 & 0.89 & 1.00 & 1.00 & 1.00 & 1.00 \\
 & Gap & 1.00 & 1.00 & 0.14 & 0.54 & 0.12 & 0.83 &  & Gap & 0.78 & 0.83 & 0.75 & 0.14 & 0.00 & 0.62 &  & Gap & 0.00 & 0.00 & 0.00 & 0.00 & 0.00 & 0.00 \\
\addlinespace[0.4pt]
S24/2 & Pred. & 4 & 2 & 7 & 11 & 14 & 10 & S30/2 & Pred. & 3 & 3 & 3 & 9 & 10 & 16 & S36/4 & Pred. & 2 & 3 & 7 & 6 & 15 & 11 \\
 & $\Delta$ & +2 & 0 & +5 & +9 & +12 & +8 &  & $\Delta$ & +1 & +1 & +1 & +7 & +8 & +14 &  & $\Delta$ & -2 & -1 & +3 & +2 & +11 & +7 \\
 & Ret. & 0.50 & 0.50 & 0.50 & 0.50 & 0.50 & 0.00 &  & Ret. & 0.00 & 0.00 & 0.50 & 0.50 & 0.00 & 1.00 &  & Ret. & 1.00 & 1.00 & 1.00 & 0.00 & 1.00 & 1.00 \\
 & Grp. & 0.73 & 0.63 & 0.66 & 0.56 & 0.58 & 0.64 &  & Grp. & 0.80 & 0.82 & 0.88 & 0.55 & 0.52 & 0.55 &  & Grp. & 0.64 & 0.56 & 0.55 & 0.51 & 0.31 & 0.36 \\
 & Ord. & 0.00 & 1.00 & 1.00 & 1.00 & 1.00 & 0.00 &  & Ord. & 0.00 & 0.00 & 1.00 & 1.00 & 0.00 & 1.00 &  & Ord. & 1.00 & 1.00 & 0.00 & 0.00 & 1.00 & 1.00 \\
 & Grd. & 0.00 & 0.33 & 0.59 & 0.56 & 0.15 & 0.00 &  & Grd. & 0.00 & 0.00 & 0.67 & 1.00 & 0.00 & 0.57 &  & Grd. & 0.15 & 0.34 & 0.83 & 0.00 & 0.76 & 0.68 \\
 & Gap & 1.00 & 0.50 & 0.00 & 0.17 & 0.86 & 0.00 &  & Gap & 0.00 & 0.00 & 0.50 & 0.00 & 0.00 & 0.50 &  & Gap & 0.00 & 0.00 & 0.25 & 0.00 & 0.00 & 0.00 \\
\bottomrule
\end{tabular}
\caption{Complete RQ1 case-level results (2/4). This vertical panel covers S19--S36. Case/GT gives the compact case ID and reference-chain length; $\Delta=\mathrm{Pred.}-\mathrm{GT}$. Metric abbreviations follow the main text.}
\label{tab:app-rq1-complete-2}
\end{table*}

\begin{table*}[t]
\centering
\scriptsize
\setlength{\tabcolsep}{1.25pt}
\begin{tabular}{@{}llcccccc@{\hspace{2pt}}llcccccc@{\hspace{2pt}}llcccccc@{}}
\toprule
Case/GT & Metric & \rotatebox{90}{Qwen-3-32b} & \rotatebox{90}{Llama4-17b-Scout} & \rotatebox{90}{DeepSeek-V4-Pro} & \rotatebox{90}{GLM-5.2} & \rotatebox{90}{GLM-5.2-T} & \rotatebox{90}{GPT-5.5} & Case/GT & Metric & \rotatebox{90}{Qwen-3-32b} & \rotatebox{90}{Llama4-17b-Scout} & \rotatebox{90}{DeepSeek-V4-Pro} & \rotatebox{90}{GLM-5.2} & \rotatebox{90}{GLM-5.2-T} & \rotatebox{90}{GPT-5.5} & Case/GT & Metric & \rotatebox{90}{Qwen-3-32b} & \rotatebox{90}{Llama4-17b-Scout} & \rotatebox{90}{DeepSeek-V4-Pro} & \rotatebox{90}{GLM-5.2} & \rotatebox{90}{GLM-5.2-T} & \rotatebox{90}{GPT-5.5} \\
\cmidrule(lr){1-8}\cmidrule(lr){9-16}\cmidrule(lr){17-24}
M01/13 & Pred. & 9 & 3 & 17 & 11 & 17 & 17 & M07/11 & Pred. & 5 & 1 & 7 & 6 & 7 & 12 & M13/11 & Pred. & 4 & 3 & 9 & 13 & 8 & 8 \\
 & $\Delta$ & -4 & -10 & +4 & -2 & +4 & +4 &  & $\Delta$ & -6 & -10 & -4 & -5 & -4 & +1 &  & $\Delta$ & -7 & -8 & -2 & +2 & -3 & -3 \\
 & Ret. & 1.00 & 1.00 & 0.85 & 0.54 & 1.00 & 1.00 &  & Ret. & 0.91 & 1.00 & 0.91 & 1.00 & 1.00 & 1.00 &  & Ret. & 0.91 & 1.00 & 0.82 & 0.91 & 1.00 & 1.00 \\
 & Grp. & 0.69 & 0.35 & 0.72 & 0.64 & 0.88 & 0.88 &  & Grp. & 0.55 & 0.31 & 0.45 & 0.63 & 0.60 & 0.55 &  & Grp. & 0.51 & 0.60 & 0.56 & 0.52 & 0.45 & 0.54 \\
 & Ord. & 1.00 & 1.00 & 0.91 & 0.95 & 0.99 & 0.99 &  & Ord. & 1.00 & 1.00 & 0.80 & 0.73 & 0.86 & 0.98 &  & Ord. & 1.00 & 0.33 & 0.96 & 0.97 & 1.00 & 1.00 \\
 & Grd. & 0.96 & 0.42 & 0.83 & 0.71 & 0.93 & 0.93 &  & Grd. & 0.48 & 1.00 & 0.25 & 0.67 & 0.62 & 0.68 &  & Grd. & 0.47 & 0.41 & 0.59 & 0.65 & 0.31 & 0.49 \\
 & Gap & 0.35 & 0.65 & 0.00 & 0.00 & 0.00 & 0.00 &  & Gap & 0.10 & 0.91 & 0.10 & 0.18 & 0.09 & 0.00 &  & Gap & 0.10 & 0.27 & 0.11 & 0.00 & 0.09 & 0.18 \\
\addlinespace[0.4pt]
M02/13 & Pred. & 10 & 3 & 12 & 12 & 10 & 22 & M08/11 & Pred. & 4 & 1 & 7 & 8 & 9 & 11 & M14/11 & Pred. & 6 & 3 & 4 & 5 & 5 & 10 \\
 & $\Delta$ & -3 & -10 & -1 & -1 & -3 & +9 &  & $\Delta$ & -7 & -10 & -4 & -3 & -2 & 0 &  & $\Delta$ & -5 & -8 & -7 & -6 & -6 & -1 \\
 & Ret. & 0.46 & 0.77 & 0.54 & 0.54 & 0.54 & 0.77 &  & Ret. & 1.00 & 1.00 & 1.00 & 1.00 & 1.00 & 1.00 &  & Ret. & 1.00 & 1.00 & 0.55 & 1.00 & 1.00 & 1.00 \\
 & Grp. & 0.62 & 0.37 & 0.63 & 0.61 & 0.56 & 0.56 &  & Grp. & 0.62 & 0.31 & 0.78 & 0.84 & 0.90 & 1.00 &  & Grp. & 0.78 & 0.53 & 0.62 & 0.71 & 0.71 & 1.00 \\
 & Ord. & 0.87 & 0.67 & 1.00 & 1.00 & 1.00 & 1.00 &  & Ord. & 1.00 & 1.00 & 0.81 & 0.93 & 0.97 & 1.00 &  & Ord. & 0.93 & 0.33 & 1.00 & 1.00 & 1.00 & 0.98 \\
 & Grd. & 0.76 & 0.50 & 0.80 & 0.73 & 0.69 & 0.75 &  & Grd. & 0.83 & 1.00 & 0.83 & 0.88 & 0.93 & 1.00 &  & Grd. & 0.83 & 0.63 & 0.92 & 0.75 & 0.75 & 1.00 \\
 & Gap & 0.20 & 0.59 & 0.16 & 0.00 & 0.00 & 0.10 &  & Gap & 0.45 & 0.91 & 0.00 & 0.00 & 0.00 & 0.00 &  & Gap & 0.18 & 0.27 & 0.17 & 0.09 & 0.09 & 0.09 \\
\addlinespace[0.4pt]
M03/13 & Pred. & 4 & 3 & 16 & 10 & 17 & 12 & M09/11 & Pred. & 3 & 1 & 12 & 9 & 22 & 17 & M15/11 & Pred. & 5 & 3 & 8 & 7 & 10 & 18 \\
 & $\Delta$ & -9 & -10 & +3 & -3 & +4 & -1 &  & $\Delta$ & -8 & -10 & +1 & -2 & +11 & +6 &  & $\Delta$ & -6 & -8 & -3 & -4 & -1 & +7 \\
 & Ret. & 0.62 & 0.62 & 0.62 & 0.62 & 0.54 & 0.54 &  & Ret. & 1.00 & 1.00 & 0.55 & 1.00 & 0.91 & 1.00 &  & Ret. & 1.00 & 1.00 & 0.91 & 0.91 & 0.91 & 1.00 \\
 & Grp. & 0.38 & 0.34 & 0.59 & 0.64 & 0.67 & 0.81 &  & Grp. & 0.53 & 0.31 & 0.57 & 0.51 & 0.42 & 0.40 &  & Grp. & 0.71 & 0.53 & 0.60 & 0.54 & 0.52 & 0.43 \\
 & Ord. & 1.00 & 1.00 & 1.00 & 1.00 & 1.00 & 0.95 &  & Ord. & 1.00 & 1.00 & 1.00 & 1.00 & 1.00 & 0.98 &  & Ord. & 1.00 & 0.33 & 0.90 & 1.00 & 0.90 & 0.90 \\
 & Grd. & 0.46 & 0.39 & 0.72 & 0.78 & 0.81 & 0.88 &  & Grd. & 0.56 & 1.00 & 0.79 & 0.51 & 0.61 & 0.53 &  & Grd. & 0.52 & 0.47 & 0.65 & 0.51 & 0.42 & 0.37 \\
 & Gap & 0.71 & 0.55 & 0.32 & 0.00 & 0.05 & 0.00 &  & Gap & 0.27 & 0.91 & 0.00 & 0.00 & 0.20 & 0.00 &  & Gap & 0.55 & 0.27 & 0.40 & 0.10 & 0.00 & 0.09 \\
\addlinespace[0.4pt]
M04/9 & Pred. & 4 & 2 & 7 & 9 & 9 & 17 & M10/9 & Pred. & 6 & 3 & 19 & 10 & 11 & 12 & M16/9 & Pred. & 7 & 3 & 12 & 9 & 13 & 17 \\
 & $\Delta$ & -5 & -7 & -2 & 0 & 0 & +8 &  & $\Delta$ & -3 & -6 & +10 & +1 & +2 & +3 &  & $\Delta$ & -2 & -6 & +3 & 0 & +4 & +8 \\
 & Ret. & 1.00 & 1.00 & 0.78 & 0.56 & 1.00 & 1.00 &  & Ret. & 1.00 & 1.00 & 1.00 & 0.67 & 0.89 & 1.00 &  & Ret. & 1.00 & 1.00 & 1.00 & 1.00 & 1.00 & 1.00 \\
 & Grp. & 0.64 & 0.51 & 0.65 & 0.56 & 0.57 & 0.36 &  & Grp. & 0.60 & 0.57 & 0.43 & 0.39 & 0.45 & 0.45 &  & Grp. & 0.38 & 0.28 & 0.51 & 0.42 & 0.56 & 0.51 \\
 & Ord. & 1.00 & 1.00 & 1.00 & 1.00 & 1.00 & 0.94 &  & Ord. & 1.00 & 1.00 & 0.80 & 1.00 & 1.00 & 0.81 &  & Ord. & 0.86 & 0.00 & 1.00 & 1.00 & 1.00 & 0.97 \\
 & Grd. & 0.54 & 0.34 & 0.77 & 0.79 & 0.68 & 0.56 &  & Grd. & 0.58 & 0.78 & 0.58 & 0.28 & 0.39 & 0.53 &  & Grd. & 0.33 & 0.43 & 0.58 & 0.36 & 0.62 & 0.56 \\
 & Gap & 0.11 & 0.56 & 0.29 & 0.00 & 0.00 & 0.00 &  & Gap & 0.22 & 0.44 & 0.33 & 0.00 & 0.00 & 0.00 &  & Gap & 0.78 & 0.56 & 0.31 & 0.44 & 0.00 & 0.42 \\
\addlinespace[0.4pt]
M05/9 & Pred. & 5 & 2 & 4 & 7 & 7 & 9 & M11/9 & Pred. & 5 & 3 & 4 & 7 & 7 & 9 & M17/9 & Pred. & 8 & 3 & 6 & 19 & 10 & 22 \\
 & $\Delta$ & -4 & -7 & -5 & -2 & -2 & 0 &  & $\Delta$ & -4 & -6 & -5 & -2 & -2 & 0 &  & $\Delta$ & -1 & -6 & -3 & +10 & +1 & +13 \\
 & Ret. & 0.89 & 1.00 & 0.44 & 1.00 & 1.00 & 1.00 &  & Ret. & 1.00 & 1.00 & 0.67 & 1.00 & 1.00 & 1.00 &  & Ret. & 0.78 & 0.44 & 0.89 & 0.89 & 0.33 & 1.00 \\
 & Grp. & 0.80 & 0.50 & 0.71 & 0.88 & 0.88 & 1.00 &  & Grp. & 0.80 & 0.62 & 0.71 & 0.88 & 0.88 & 1.00 &  & Grp. & 0.50 & 0.49 & 0.50 & 0.56 & 0.50 & 0.48 \\
 & Ord. & 1.00 & 1.00 & 1.00 & 0.95 & 0.95 & 0.94 &  & Ord. & 1.00 & 1.00 & 1.00 & 1.00 & 1.00 & 0.86 &  & Ord. & 1.00 & 0.00 & 1.00 & 0.96 & 1.00 & 1.00 \\
 & Grd. & 0.90 & 0.67 & 1.00 & 0.90 & 0.90 & 1.00 &  & Grd. & 0.88 & 0.83 & 0.92 & 0.93 & 0.90 & 1.00 &  & Grd. & 0.44 & 0.00 & 0.44 & 0.45 & 0.33 & 0.50 \\
 & Gap & 0.12 & 0.56 & 0.00 & 0.00 & 0.00 & 0.00 &  & Gap & 0.11 & 0.44 & 0.17 & 0.00 & 0.00 & 0.00 &  & Gap & 0.29 & 1.00 & 0.56 & 0.00 & 0.88 & 0.36 \\
\addlinespace[0.4pt]
M06/9 & Pred. & 4 & 2 & 14 & 7 & 7 & 17 & M12/9 & Pred. & 3 & 3 & 7 & 8 & 8 & 13 & M18/9 & Pred. & 4 & 3 & 11 & 11 & 10 & 13 \\
 & $\Delta$ & -5 & -7 & +5 & -2 & -2 & +8 &  & $\Delta$ & -6 & -6 & -2 & -1 & -1 & +4 &  & $\Delta$ & -5 & -6 & +2 & +2 & +1 & +4 \\
 & Ret. & 1.00 & 1.00 & 1.00 & 1.00 & 0.78 & 1.00 &  & Ret. & 0.89 & 1.00 & 1.00 & 0.22 & 0.89 & 1.00 &  & Ret. & 0.56 & 0.44 & 1.00 & 1.00 & 1.00 & 1.00 \\
 & Grp. & 0.62 & 0.52 & 0.55 & 0.54 & 0.52 & 0.35 &  & Grp. & 0.59 & 0.54 & 0.54 & 0.51 & 0.50 & 0.44 &  & Grp. & 0.37 & 0.46 & 0.48 & 0.49 & 0.48 & 0.46 \\
 & Ord. & 0.33 & 1.00 & 0.93 & 1.00 & 0.83 & 0.94 &  & Ord. & 1.00 & 1.00 & 1.00 & 1.00 & 1.00 & 0.67 &  & Ord. & 1.00 & 0.00 & 0.71 & 1.00 & 0.86 & 1.00 \\
 & Grd. & 0.72 & 0.31 & 0.73 & 0.45 & 0.49 & 0.53 &  & Grd. & 0.50 & 0.44 & 0.57 & 0.40 & 0.45 & 0.46 &  & Grd. & 0.33 & 0.45 & 0.36 & 0.44 & 0.44 & 0.47 \\
 & Gap & 0.33 & 0.56 & 0.11 & 0.11 & 0.00 & 0.00 &  & Gap & 0.88 & 0.44 & 0.22 & 0.00 & 0.00 & 0.00 &  & Gap & 0.71 & 0.50 & 0.50 & 0.29 & 0.28 & 0.30 \\
\bottomrule
\end{tabular}
\caption{Complete RQ1 case-level results (3/4). This vertical panel covers M01--M18. Case/GT gives the compact case ID and reference-chain length; $\Delta=\mathrm{Pred.}-\mathrm{GT}$. Metric abbreviations follow the main text.}
\label{tab:app-rq1-complete-3}
\end{table*}

\begin{table*}[t]
\centering
\scriptsize
\setlength{\tabcolsep}{1.25pt}
\begin{tabular}{@{}llcccccc@{\hspace{2pt}}llcccccc@{\hspace{2pt}}llcccccc@{}}
\toprule
Case/GT & Metric & \rotatebox{90}{Qwen-3-32b} & \rotatebox{90}{Llama4-17b-Scout} & \rotatebox{90}{DeepSeek-V4-Pro} & \rotatebox{90}{GLM-5.2} & \rotatebox{90}{GLM-5.2-T} & \rotatebox{90}{GPT-5.5} & Case/GT & Metric & \rotatebox{90}{Qwen-3-32b} & \rotatebox{90}{Llama4-17b-Scout} & \rotatebox{90}{DeepSeek-V4-Pro} & \rotatebox{90}{GLM-5.2} & \rotatebox{90}{GLM-5.2-T} & \rotatebox{90}{GPT-5.5} & Case/GT & Metric & \rotatebox{90}{Qwen-3-32b} & \rotatebox{90}{Llama4-17b-Scout} & \rotatebox{90}{DeepSeek-V4-Pro} & \rotatebox{90}{GLM-5.2} & \rotatebox{90}{GLM-5.2-T} & \rotatebox{90}{GPT-5.5} \\
\cmidrule(lr){1-8}\cmidrule(lr){9-16}\cmidrule(lr){17-24}
L01/47 & Pred. & 7 & 3 & 16 & 18 & 22 & 29 & L07/27 & Pred. & 3 & 3 & 7 & 13 & 28 & 25 & L13/24 & Pred. & 6 & 1 & 20 & 17 & 21 & 26 \\
 & $\Delta$ & -40 & -44 & -31 & -29 & -25 & -18 &  & $\Delta$ & -24 & -24 & -20 & -14 & +1 & -2 &  & $\Delta$ & -18 & -23 & -4 & -7 & -3 & +2 \\
 & Ret. & 0.72 & 0.89 & 0.98 & 0.62 & 0.72 & 0.83 &  & Ret. & 0.78 & 0.96 & 0.22 & 0.44 & 0.96 & 0.78 &  & Ret. & 0.92 & 1.00 & 1.00 & 0.92 & 1.00 & 1.00 \\
 & Grp. & 0.23 & 0.16 & 0.48 & 0.56 & 0.60 & 0.62 &  & Grp. & 0.32 & 0.26 & 0.44 & 0.58 & 0.53 & 0.59 &  & Grp. & 0.38 & 0.16 & 0.70 & 0.66 & 0.69 & 0.71 \\
 & Ord. & 0.40 & 1.00 & 0.85 & 0.91 & 0.93 & 0.97 &  & Ord. & 1.00 & 0.67 & 1.00 & 0.82 & 0.86 & 1.00 &  & Ord. & 0.80 & 1.00 & 0.70 & 0.64 & 0.62 & 0.67 \\
 & Grd. & 0.39 & 0.47 & 0.82 & 0.81 & 0.82 & 0.79 &  & Grd. & 0.38 & 0.54 & 0.52 & 0.75 & 0.67 & 0.77 &  & Grd. & 0.53 & 0.29 & 0.76 & 0.73 & 0.73 & 0.75 \\
 & Gap & 0.41 & 0.74 & 0.52 & 0.00 & 0.21 & 0.13 &  & Gap & 0.86 & 0.54 & 0.50 & 0.00 & 0.04 & 0.00 &  & Gap & 0.52 & 0.82 & 0.00 & 0.00 & 0.00 & 0.00 \\
\addlinespace[0.4pt]
L02/47 & Pred. & 5 & 3 & 18 & 22 & 29 & 40 & L08/27 & Pred. & 4 & 3 & 26 & 21 & 22 & 27 & L14/24 & Pred. & 2 & 2 & 21 & 16 & 14 & 6 \\
 & $\Delta$ & -42 & -44 & -29 & -25 & -18 & -7 &  & $\Delta$ & -23 & -24 & -1 & -6 & -5 & 0 &  & $\Delta$ & -22 & -22 & -3 & -8 & -10 & -18 \\
 & Ret. & 0.83 & 0.91 & 0.68 & 0.68 & 1.00 & 1.00 &  & Ret. & 1.00 & 1.00 & 1.00 & 0.81 & 1.00 & 1.00 &  & Ret. & 0.79 & 0.67 & 0.79 & 0.79 & 0.83 & 0.96 \\
 & Grp. & 0.16 & 0.16 & 0.58 & 0.66 & 0.76 & 0.93 &  & Grp. & 0.31 & 0.26 & 0.94 & 0.90 & 0.90 & 1.00 &  & Grp. & 0.19 & 0.20 & 0.59 & 0.44 & 0.51 & 0.41 \\
 & Ord. & 1.00 & 0.67 & 0.95 & 0.83 & 0.87 & 0.95 &  & Ord. & 0.83 & 0.67 & 0.88 & 0.80 & 1.00 & 1.00 &  & Ord. & 0.00 & 1.00 & 0.70 & 0.80 & 0.86 & 1.00 \\
 & Grd. & 0.34 & 0.44 & 0.89 & 0.89 & 0.84 & 0.95 &  & Grd. & 0.52 & 0.37 & 0.97 & 0.98 & 0.92 & 1.00 &  & Grd. & 0.37 & 0.40 & 0.73 & 0.40 & 0.59 & 0.46 \\
 & Gap & 0.28 & 0.72 & 0.25 & 0.06 & 0.00 & 0.02 &  & Gap & 0.56 & 0.48 & 0.07 & 0.00 & 0.00 & 0.00 &  & Gap & 0.70 & 0.75 & 0.00 & 0.04 & 0.07 & 0.10 \\
\addlinespace[0.4pt]
L03/47 & Pred. & 6 & 3 & 16 & 18 & 24 & 45 & L09/27 & Pred. & 9 & 3 & 27 & 22 & 24 & 33 & L15/24 & Pred. & 3 & 2 & 10 & 16 & 20 & 5 \\
 & $\Delta$ & -41 & -44 & -31 & -29 & -23 & -2 &  & $\Delta$ & -18 & -24 & 0 & -5 & -3 & +6 &  & $\Delta$ & -21 & -22 & -14 & -8 & -4 & -19 \\
 & Ret. & 0.89 & 0.87 & 0.45 & 0.51 & 0.96 & 0.89 &  & Ret. & 0.70 & 1.00 & 0.78 & 0.74 & 0.85 & 0.85 &  & Ret. & 0.17 & 0.21 & 0.67 & 0.67 & 0.54 & 0.62 \\
 & Grp. & 0.25 & 0.16 & 0.47 & 0.53 & 0.59 & 0.71 &  & Grp. & 0.43 & 0.26 & 0.76 & 0.76 & 0.60 & 0.58 &  & Grp. & 0.26 & 0.24 & 0.50 & 0.52 & 0.51 & 0.33 \\
 & Ord. & 0.60 & 1.00 & 0.65 & 0.75 & 0.89 & 1.00 &  & Ord. & 1.00 & 0.67 & 0.73 & 0.90 & 1.00 & 0.97 &  & Ord. & 0.00 & 1.00 & 0.67 & 0.60 & 0.83 & 1.00 \\
 & Grd. & 0.61 & 0.58 & 0.81 & 0.85 & 0.64 & 0.86 &  & Grd. & 1.00 & 0.43 & 0.97 & 0.93 & 0.75 & 0.84 &  & Grd. & 0.00 & 0.29 & 0.34 & 0.38 & 0.47 & 0.67 \\
 & Gap & 0.64 & 0.76 & 0.19 & 0.08 & 0.09 & 0.02 &  & Gap & 0.89 & 0.52 & 0.00 & 0.00 & 0.00 & 0.00 &  & Gap & 1.00 & 0.50 & 0.22 & 0.12 & 0.38 & 0.90 \\
\addlinespace[0.4pt]
L04/68 & Pred. & 3 & 3 & 22 & 33 & 28 & 45 & L10/26 & Pred. & 3 & 2 & 7 & 18 & 18 & 27 &  &  &  &  &  &  &  &  \\
 & $\Delta$ & -65 & -65 & -46 & -35 & -40 & -23 &  & $\Delta$ & -23 & -24 & -19 & -8 & -8 & +1 &  &  &  &  &  &  &  &  \\
 & Ret. & 0.78 & 0.69 & 0.47 & 0.44 & 0.47 & 0.72 &  & Ret. & 0.81 & 0.96 & 0.92 & 1.00 & 1.00 & 1.00 &  &  &  &  &  &  &  &  \\
 & Grp. & 0.08 & 0.11 & 0.51 & 0.51 & 0.60 & 0.81 &  & Grp. & 0.21 & 0.19 & 0.29 & 0.59 & 0.55 & 0.69 &  &  &  &  &  &  &  &  \\
 & Ord. & 0.00 & 0.33 & 0.90 & 0.89 & 0.96 & 1.00 &  & Ord. & 0.00 & 0.00 & 0.90 & 0.91 & 0.77 & 0.89 &  &  &  &  &  &  &  &  \\
 & Grd. & 0.34 & 0.30 & 0.92 & 0.90 & 1.00 & 0.97 &  & Grd. & 0.65 & 0.57 & 0.47 & 0.65 & 0.62 & 0.74 &  &  &  &  &  &  &  &  \\
 & Gap & 0.81 & 0.64 & 0.09 & 0.00 & 0.12 & 0.00 &  & Gap & 0.88 & 0.81 & 0.57 & 0.10 & 0.29 & 0.09 &  &  &  &  &  &  &  &  \\
\addlinespace[0.4pt]
L05/68 & Pred. & 6 & 3 & 34 & 27 & 3 & 43 & L11/26 & Pred. & 2 & 2 & 6 & 25 & 21 & 12 &  &  &  &  &  &  &  &  \\
 & $\Delta$ & -62 & -65 & -34 & -41 & -65 & -25 &  & $\Delta$ & -24 & -24 & -20 & -1 & -5 & -14 &  &  &  &  &  &  &  &  \\
 & Ret. & 0.75 & 0.69 & 0.82 & 0.47 & 0.85 & 0.68 &  & Ret. & 0.77 & 0.46 & 0.81 & 0.96 & 0.85 & 0.96 &  &  &  &  &  &  &  &  \\
 & Grp. & 0.19 & 0.11 & 0.68 & 0.58 & 0.11 & 0.73 &  & Grp. & 0.24 & 0.30 & 0.31 & 0.52 & 0.46 & 0.49 &  &  &  &  &  &  &  &  \\
 & Ord. & 1.00 & 0.67 & 0.75 & 0.95 & 1.00 & 0.98 &  & Ord. & 1.00 & 1.00 & 0.93 & 0.87 & 0.87 & 0.86 &  &  &  &  &  &  &  &  \\
 & Grd. & 0.72 & 0.29 & 0.92 & 0.94 & 0.63 & 0.96 &  & Grd. & 0.25 & 0.86 & 0.57 & 0.56 & 0.50 & 0.68 &  &  &  &  &  &  &  &  \\
 & Gap & 0.73 & 0.62 & 0.16 & 0.00 & 0.81 & 0.09 &  & Gap & 0.88 & 0.60 & 0.57 & 0.00 & 0.00 & 0.54 &  &  &  &  &  &  &  &  \\
\addlinespace[0.4pt]
L06/68 & Pred. & 3 & 3 & 24 & 28 & 47 & 44 & L12/26 & Pred. & 3 & 2 & 11 & 8 & 17 & 15 &  &  &  &  &  &  &  &  \\
 & $\Delta$ & -65 & -65 & -44 & -40 & -21 & -24 &  & $\Delta$ & -23 & -24 & -15 & -18 & -9 & -11 &  &  &  &  &  &  &  &  \\
 & Ret. & 0.65 & 0.76 & 0.60 & 0.47 & 0.82 & 0.62 &  & Ret. & 0.42 & 0.38 & 0.15 & 0.23 & 0.65 & 0.62 &  &  &  &  &  &  &  &  \\
 & Grp. & 0.11 & 0.11 & 0.40 & 0.60 & 0.74 & 0.62 &  & Grp. & 0.17 & 0.24 & 0.47 & 0.54 & 0.50 & 0.43 &  &  &  &  &  &  &  &  \\
 & Ord. & 0.67 & 0.33 & 0.77 & 0.88 & 0.91 & 0.93 &  & Ord. & 1.00 & 1.00 & 1.00 & 1.00 & 0.86 & 0.73 &  &  &  &  &  &  &  &  \\
 & Grd. & 0.46 & 0.30 & 0.69 & 0.96 & 0.89 & 0.87 &  & Grd. & 0.50 & 0.50 & 0.11 & 0.30 & 0.53 & 0.63 &  &  &  &  &  &  &  &  \\
 & Gap & 0.68 & 0.67 & 0.10 & 0.03 & 0.05 & 0.21 &  & Gap & 0.88 & 0.88 & 0.00 & 0.00 & 0.00 & 0.09 &  &  &  &  &  &  &  &  \\
\bottomrule
\end{tabular}
\caption{Complete RQ1 case-level results (4/4). This vertical panel covers L01--L15. Case/GT gives the compact case ID and reference-chain length; $\Delta=\mathrm{Pred.}-\mathrm{GT}$. Metric abbreviations follow the main text.}
\label{tab:app-rq1-complete-4}
\end{table*}

\endgroup

\subsection{RQ2: Progressive Failure Anatomy}Table~\ref{tab:app-rq2-funnel} reports all RQ2 first-failure counts. Each coverable reference step is assigned once to E1--E4/OK. The overall, evidence-profile, and chain-length rows use the denominators shown in \(N\); every stage cell gives the count followed by its row percentage.

\subsection{RQ4: Budget Sensitivity}
Table~\ref{tab:app-rq4-budget} reports the complete GLM-5.2 budget sweeps on the same R24 cases used in the main paper.

\subsection{Auxiliary Evidence-Dependence Control}
\label{app:rq5-control}

Because these source incidents are public, a model could recall an attack chain instead of reconstructing it from supplied evidence. We therefore evaluate the frozen PUBLIC-15 clean subset---13 AutoLabel cases and the two OTRF APT29 days---under two conditions. C0 supplies only an anonymous minimal brief, with no evidence cards, retrieval documents, or graph, and disables grounding checks. C1 retains the evidence environment but applies a deterministic, type-preserving bijection to identifiers such as CVEs, hosts, IP addresses, domains, paths, and components across the brief, evidence, retrieval corpus, graph, and reference chain; evidence IDs remain fixed and grounding checks remain enabled. The mutation audit reports zero leaked original tokens and zero evidence-ID preservation violations.

Figure~\ref{fig:app-rq5-conditions} includes Original only to show what C0 removes and C1 anonymizes; the paired analysis uses C0 and C1.

This five-model auxiliary bundle uses an eight-turn ceiling and top-$k$ 32. It predates the final 15-turn protocol and is not pooled with RQ1/RQ2.

Tables~\ref{tab:app-rq5-control} and~\ref{tab:app-rq5-recovery} report descriptive scores and paired differences. C0 Ret. and Ord. are valid zeros. Grp. is N/A because empty predictions inherit the evaluator's background-cluster default, Grd. because grounding is disabled, and Gap because C0 has no observed evidence. Paired differences therefore report only structural F1, pairwise order, and Ret.

\begin{table*}[!t]
\centering
\small
\setlength{\tabcolsep}{4.2pt}
\begin{tabular}{lrrrrrrr}
\toprule
Model (cond.) & Nonempty & Struct. F1 & Ret. & Grp. & Ord. & Grd. & Gap \\
\midrule
Qwen2.5-7B (C0) & 26.7 & 0.0 & 0.0 & N/A & 0.0 & N/A & N/A \\ Qwen2.5-7B (C1) & 100.0 & 56.3 & 97.7 & 61.8 & 100.0 & 71.8 & 33.8 \\
GLM-5.2 (C0) & 0.0 & 0.0 & 0.0 & N/A & 0.0 & N/A & N/A \\ GLM-5.2 (C1) & 100.0 & 59.9 & 100.0 & 62.6 & 96.5 & 66.5 & 9.3 \\
Llama2-7B (C0) & 0.0 & 0.0 & 0.0 & N/A & 0.0 & N/A & N/A \\ Llama2-7B (C1) & 93.3 & 16.9 & 99.7 & 61.9 & 84.4 & 60.4 & 56.7 \\
GPT-5.5 (C0) & 0.0 & 0.0 & 0.0 & N/A & 0.0 & N/A & N/A \\ GPT-5.5 (C1) & 100.0 & 53.7 & 100.0 & 65.7 & 96.0 & 64.6 & 21.5 \\
DeepSeek-V4-Pro (C0) & 0.0 & 0.0 & 0.0 & N/A & 0.0 & N/A & N/A \\ DeepSeek-V4-Pro (C1) & 100.0 & 53.2 & 98.9 & 66.5 & 97.0 & 68.5 & 24.2 \\
\bottomrule
\end{tabular}
\caption{Complete protocol-tagged auxiliary C0/C1 results (\%; $n=15$ cases per model and condition), using an eight-turn ceiling and top-$k$ 32. Both conditions use the same columns. C0 Grp., Grd., and Gap are N/A for the protocol reasons stated in the text.}
\label{tab:app-rq5-control}
\end{table*}

Table~\ref{tab:app-rq5-recovery} gives 95\% case-bootstrap intervals for the three eligible paired differences. All five models have positive C1--C0 intervals for structural F1, pairwise order, and retrieval coverage, while C0 structural F1 and order are zero. This supports dependence on supplied evidence within this auxiliary protocol. Without a matched Original condition, it does not estimate the effect of anonymization or establish absence of training contamination.

\begin{table*}[!t]
\centering
\small
\setlength{\tabcolsep}{3.5pt}
\begin{tabular}{lccc}
\toprule
Model & Structural F1 & Pairwise order & Ret. \\
\midrule
Qwen2.5-7B & 56.3 [36.8, 74.9] & 80.0 [60.0, 100.0] & 97.7 [94.4, 100.0] \\
GLM-5.2 & 59.9 [47.4, 72.7] & 96.4 [90.7, 100.0] & 100.0 [100.0, 100.0] \\
Llama2-7B & 16.9 [6.0, 29.6] & 40.0 [13.3, 66.7] & 99.7 [99.2, 100.0] \\
GPT-5.5 & 53.7 [40.7, 67.6] & 91.6 [77.1, 99.8] & 100.0 [100.0, 100.0] \\
DeepSeek-V4-Pro & 53.2 [42.5, 63.7] & 95.3 [88.6, 100.0] & 98.9 [97.3, 100.0] \\
\bottomrule
\end{tabular}
\caption{Protocol-tagged auxiliary C1--C0 differences in percentage points with 95\% case-bootstrap intervals ($n=15$; 10,000 resamples; seed 20,260,711). Only metrics with valid C0 baselines are reported; Grp., Grd., and Gap are excluded for the protocol reasons stated in the text.}
\label{tab:app-rq5-recovery}
\end{table*}

\begin{table*}[!t]\centering\scriptsize\setlength{\tabcolsep}{2.5pt}\renewcommand{\arraystretch}{0.94}\begin{tabular}{llr*{5}{r}}\toprule Model & Slice & \(N\) & E1 & E2 & E3 & E4 & OK \\\midrule Qwen-3-32b & Overall & 849 & 190 (22.4\%) & 333 (39.2\%) & 74 (8.7\%) & 114 (13.4\%) & 138 (16.3\%) \\ & Profile: clean & 283 & 33 (11.7\%) & 99 (35.0\%) & 39 (13.8\%) & 48 (17.0\%) & 64 (22.6\%) \\ & Profile: noisy & 283 & 59 (20.8\%) & 120 (42.4\%) & 22 (7.8\%) & 32 (11.3\%) & 50 (17.7\%) \\ & Profile: raw & 283 & 98 (34.6\%) & 114 (40.3\%) & 13 (4.6\%) & 34 (12.0\%) & 24 (8.5\%) \\ & Length: S & 87 & 19 (21.8\%) & 17 (19.5\%) & 27 (31.0\%) & 2 (2.3\%) & 22 (25.3\%) \\ & Length: M & 186 & 22 (11.8\%) & 47 (25.3\%) & 19 (10.2\%) & 19 (10.2\%) & 79 (42.5\%) \\ & Length: L & 576 & 149 (25.9\%) & 269 (46.7\%) & 28 (4.9\%) & 93 (16.1\%) & 37 (6.4\%) \\\midrule DeepSeek-V4-Pro & Overall & 849 & 234 (27.6\%) & 99 (11.7\%) & 64 (7.5\%) & 280 (33.0\%) & 172 (20.3\%) \\ & Profile: clean & 283 & 44 (15.5\%) & 25 (8.8\%) & 26 (9.2\%) & 113 (39.9\%) & 75 (26.5\%) \\ & Profile: noisy & 283 & 81 (28.6\%) & 54 (19.1\%) & 24 (8.5\%) & 71 (25.1\%) & 53 (18.7\%) \\ & Profile: raw & 283 & 109 (38.5\%) & 20 (7.1\%) & 14 (4.9\%) & 96 (33.9\%) & 44 (15.5\%) \\ & Length: S & 87 & 10 (11.5\%) & 7 (8.0\%) & 24 (27.6\%) & 3 (3.4\%) & 43 (49.4\%) \\ & Length: M & 186 & 38 (20.4\%) & 24 (12.9\%) & 12 (6.5\%) & 29 (15.6\%) & 83 (44.6\%) \\ & Length: L & 576 & 186 (32.3\%) & 68 (11.8\%) & 28 (4.9\%) & 248 (43.1\%) & 46 (8.0\%) \\\midrule GLM-5.2 & Overall & 849 & 278 (32.7\%) & 13 (1.5\%) & 34 (4.0\%) & 309 (36.4\%) & 215 (25.3\%) \\ & Profile: clean & 283 & 64 (22.6\%) & 3 (1.1\%) & 16 (5.7\%) & 120 (42.4\%) & 80 (28.3\%) \\ & Profile: noisy & 283 & 96 (33.9\%) & 2 (0.7\%) & 9 (3.2\%) & 108 (38.2\%) & 68 (24.0\%) \\ & Profile: raw & 283 & 118 (41.7\%) & 8 (2.8\%) & 9 (3.2\%) & 81 (28.6\%) & 67 (23.7\%) \\ & Length: S & 87 & 15 (17.2\%) & 0 (0.0\%) & 20 (23.0\%) & 0 (0.0\%) & 52 (59.8\%) \\ & Length: M & 186 & 34 (18.3\%) & 6 (3.2\%) & 5 (2.7\%) & 36 (19.4\%) & 105 (56.5\%) \\ & Length: L & 576 & 229 (39.8\%) & 7 (1.2\%) & 9 (1.6\%) & 273 (47.4\%) & 58 (10.1\%) \\\midrule GLM-5.2-T & Overall & 849 & 141 (16.6\%) & 83 (9.8\%) & 40 (4.7\%) & 346 (40.8\%) & 239 (28.2\%) \\ & Profile: clean & 283 & 10 (3.5\%) & 49 (17.3\%) & 20 (7.1\%) & 126 (44.5\%) & 78 (27.6\%) \\ & Profile: noisy & 283 & 73 (25.8\%) & 18 (6.4\%) & 10 (3.5\%) & 107 (37.8\%) & 75 (26.5\%) \\ & Profile: raw & 283 & 58 (20.5\%) & 16 (5.7\%) & 10 (3.5\%) & 113 (39.9\%) & 86 (30.4\%) \\ & Length: S & 87 & 11 (12.6\%) & 1 (1.1\%) & 22 (25.3\%) & 0 (0.0\%) & 53 (60.9\%) \\ & Length: M & 186 & 24 (12.9\%) & 9 (4.8\%) & 1 (0.5\%) & 45 (24.2\%) & 107 (57.5\%) \\ & Length: L & 576 & 106 (18.4\%) & 73 (12.7\%) & 17 (3.0\%) & 301 (52.3\%) & 79 (13.7\%) \\\midrule GPT-5.5 & Overall & 849 & 126 (14.8\%) & 49 (5.8\%) & 78 (9.2\%) & 260 (30.6\%) & 336 (39.6\%) \\ & Profile: clean & 283 & 22 (7.8\%) & 6 (2.1\%) & 33 (11.7\%) & 114 (40.3\%) & 108 (38.2\%) \\ & Profile: noisy & 283 & 40 (14.1\%) & 17 (6.0\%) & 29 (10.2\%) & 79 (27.9\%) & 118 (41.7\%) \\ & Profile: raw & 283 & 64 (22.6\%) & 26 (9.2\%) & 16 (5.7\%) & 67 (23.7\%) & 110 (38.9\%) \\ & Length: S & 87 & 6 (6.9\%) & 2 (2.3\%) & 42 (48.3\%) & 0 (0.0\%) & 37 (42.5\%) \\ & Length: M & 186 & 9 (4.8\%) & 6 (3.2\%) & 16 (8.6\%) & 42 (22.6\%) & 113 (60.8\%) \\ & Length: L & 576 & 111 (19.3\%) & 41 (7.1\%) & 20 (3.5\%) & 218 (37.8\%) & 186 (32.3\%) \\\midrule Llama4-17b-Scout & Overall & 849 & 167 (19.7\%) & 370 (43.6\%) & 61 (7.2\%) & 153 (18.0\%) & 98 (11.5\%) \\ & Profile: clean & 283 & 26 (9.2\%) & 130 (45.9\%) & 37 (13.1\%) & 55 (19.4\%) & 35 (12.4\%) \\ & Profile: noisy & 283 & 60 (21.2\%) & 121 (42.8\%) & 18 (6.4\%) & 50 (17.7\%) & 34 (12.0\%) \\ & Profile: raw & 283 & 81 (28.6\%) & 119 (42.0\%) & 6 (2.1\%) & 48 (17.0\%) & 29 (10.2\%) \\ & Length: S & 87 & 17 (19.5\%) & 18 (20.7\%) & 21 (24.1\%) & 9 (10.3\%) & 22 (25.3\%) \\ & Length: M & 186 & 18 (9.7\%) & 87 (46.8\%) & 15 (8.1\%) & 26 (14.0\%) & 40 (21.5\%) \\ & Length: L & 576 & 132 (22.9\%) & 265 (46.0\%) & 25 (4.3\%) & 118 (20.5\%) & 36 (6.2\%) \\\bottomrule\end{tabular}\caption{Complete RQ2 progressive first-failure results. Cells report count (row percentage); \(N\) is the number of coverable reference steps. Overall rows pool all 69 cases per model; profile and length rows use the indicated strata. E1--E4/OK follow Table~\ref{tab:app-funnel-gates}.}\label{tab:app-rq2-funnel}\label{tab:app-rq2-tier}\label{tab:app-rq2-length}\end{table*}

\begin{table*}[!t]\centering\small\setlength{\tabcolsep}{3.5pt}\begin{tabular}{llrrrrrrrr}\toprule{}Sweep & Setting & Ret. & Grp. & Ord. & Grd. & Gap & Q & Tok. (K) & Lat. (s) \\\midrule{}Top-$k$ & 1  & 24.9 & 40.9 & 82.6 & 52.0 & 0.0  & 6.96 & 58.6  & 106.4 \\ & 4  & 36.2 & 47.2 & 80.8 & 54.0 & 3.1  & 5.29 & 70.5  & 142.6 \\ & 8  & 53.2 & 51.8 & 91.8 & 56.8 & 11.5 & 4.00 & 69.7  & 128.1 \\ & 16 & 58.8 & 55.5 & 92.1 & 56.8 & 9.1  & 3.17 & 78.0  & 139.5 \\ & 32 & 73.6 & 58.6 & 94.4 & 64.3 & 11.6 & 3.29 & 92.9  & 141.8 \\ & 48 & 76.0 & 62.2 & 89.7 & 63.1 & 14.4 & 3.67 & 122.0 & 192.1 \\ & 64 & 84.7 & 63.0 & 94.9 & 61.9 & 21.4 & 3.12 & 119.3 & 235.4 \\\midrule{}Turn ceiling & 3  & 53.3 & 57.7 & 86.6 & 58.2 & 11.9 & 1.25 & 47.1  & 99.5 \\ & 5  & 65.0 & 59.9 & 87.7 & 62.5 & 6.0  & 2.04 & 67.4  & 121.5 \\ & 10 & 73.0 & 57.4 & 94.9 & 61.5 & 12.1 & 2.71 & 82.2  & 135.7 \\ & 15 & 73.6 & 58.6 & 94.4 & 64.3 & 11.6 & 3.29 & 92.9  & 141.8 \\ & 20 & 76.9 & 59.4 & 87.5 & 65.8 & 11.0 & 3.21 & 92.6  & 150.1 \\ & 25 & 82.4 & 61.5 & 88.8 & 65.3 & 15.2 & 4.29 & 130.4 & 200.0 \\\bottomrule\end{tabular}\caption{Complete RQ4 GLM-5.2 budget sweeps on R24. The top-$k$ rows use a 15-turn ceiling; the turn-ceiling rows use $k=32$. Quality columns are percentages; Q is mean queries, and Tok. is mean total tokens in thousands.}\label{tab:app-rq4-budget}\label{tab:app-rq4-topk}\label{tab:app-rq4-turns}\end{table*}

\begin{figure*}[!t]
\centering
\begin{lstlisting}[style=appprompt,firstline=1,lastline=28]
You are a cybersecurity analyst reconstructing an attack chain from evidence.

Use only the incident input JSON provided by the caller. The input contains log evidence cards only; it does not contain source code, attack scripts, gold step labels, or a gold-derived ATT&CK technique catalog.

Your task:
1. Select evidence cards that are attack-relevant.
2. Group related observations into temporally ordered attack steps.
3. Assign each step a MITRE ATT&CK tactic and technique.
4. Cite supporting evidence IDs for every step.
5. Predict step-to-step edges only when the evidence or sequence supports them.

Granularity requirements:
- Reconstruct attacker actions, not broad campaign phases.
- One attack step should have exactly one primary action and one primary security intent.
- Do not merge multiple distinct commands, expressions, file targets, or ATT&CK intents into one step.
- If a step description would need words like "and then", "various", "multiple", "including", or "leading to", split it into smaller steps.
- Treat each distinct attacker-controlled request, command, resource target, or security intent as a separate candidate step when evidence supports it.
- Repeated log lines for the same action should be grouped into one step, but log lines showing different commands or different file targets should become different steps.
- A broad step is insufficient when the evidence supports several separate attacker actions or intents.
- Prefer a complete fine-grained chain over a short high-level summary. It is acceptable to output many steps when the evidence supports many distinct actions.

Coverage discipline:
- Before writing the final JSON, silently build an internal anchor coverage ledger from the visible evidence.
- For every attack-relevant evidence anchor, decide whether it creates a new step, supports an existing step, is a duplicate of an existing step, or is background.
- Do not output this ledger, but use it to avoid dropping observed attacker actions.
- If an observed anchor has a distinct action, target, host, account, file, command, URL, or security intent, prefer adding a low-confidence supported step over merging it into a broad summary.
- If many observed anchors remain uncited, review whether they are true duplicates/background or whether the chain is missing steps.
\end{lstlisting}
\caption{Base attack-chain reconstruction prompt (part 1 of 2): task, granularity, and visible-anchor coverage discipline. Text is reproduced from the frozen prompt artifact.}
\label{fig:app-prompt-base-a}
\end{figure*}

\begin{figure*}[p]
\centering
\begin{lstlisting}[style=appprompt,firstline=29,lastline=71]
Return JSON only, following this shape:

```json
{
  "incident_summary": "...",
  "attack_chain": [
    {
      "step_id": "S1",
      "order": 1,
      "attack_step": "short step title",
      "step_description": "evidence-grounded explanation",
      "tactic": "MITRE ATT&CK tactic inferred from evidence",
      "technique_id": "MITRE ATT&CK technique ID inferred from evidence",
      "technique_name": "MITRE ATT&CK technique name inferred from evidence",
      "entities": {
        "files": [],
        "processes": [],
        "hosts": [],
        "urls": [],
        "accounts": []
      },
      "supporting_evidence": ["EVIDENCE_ID"],
      "confidence": "low|medium|high"
    }
  ],
  "causal_edges": [
    {
      "source": "S1",
      "target": "S2",
      "relation": "temporal_after|enables|uses_output_of|same_session_sequence",
      "supporting_evidence": ["EVIDENCE_ID"]
    }
  ]
}
```

Rules:
- Do not cite evidence IDs that are not present in `evidence_cards`.
- Do not output a step without at least one supporting evidence ID unless you mark confidence as `low`.
- If the evidence does not support a claim, omit the claim.
- Do not reveal chain labels from any external source; infer them from evidence only.
- Before returning, check every step: if its supporting evidence contains more than one distinct command expression or file target, split the step.
- Before returning, check coverage: every observed attack-relevant anchor should either be cited by a step or intentionally treated as duplicate/background.
\end{lstlisting}
\caption{Base attack-chain reconstruction prompt (part 2 of 2): output JSON schema and evidence-grounding rules.}
\label{fig:app-prompt-base-b}
\end{figure*}

\begin{figure*}[p]
\centering
\begin{lstlisting}[style=appprompt,firstline=1,lastline=63]
Important granularity rule:
- Reconstruct fine-grained attack steps. Do not merge distinct actions into one broad phase.
- Separate distinct observed requests, commands, resource targets, and security intents when evidence supports them.
- Every output step should cite at least one evidence ID from the provided context.
- Anti-merge check: if a proposed step cites evidence for several different commands or file targets, split it before returning.
- Use sequential step IDs (S1, S2, ...) and ensure each causal edge connects adjacent or clearly related fine-grained steps.

Granularity calibration profile: forensic action units

Use this profile to avoid broad narrative summaries. The goal is to make the
predicted chain use the same kind of action-unit granularity that an incident
responder would use when reading logs.

Do not infer or target any fixed number of steps. The number of steps must come
only from the evidence.

Silently maintain an anchor coverage ledger while reading evidence:
- anchor: the visible request, command, alert, file access, network action,
  account use, host change, URL access, or other concrete attacker behavior;
- decision: new step, supporting evidence for an existing step, duplicate, or
  background;
- reason: the action/target/intent evidence that justifies the decision.

Do not output this ledger. Use it only to prevent long chains from being
compressed into a few narrative phases.

Split a candidate step when the cited evidence shows a different:
- observed request, command, or interpreter expression;
- resource target, such as a different file, directory, URL, account, token, key,
  host, or dataset;
- security intent, such as authentication, code execution, discovery, credential
  access, collection, or exfiltration;
- dependency role, such as using one observation to decide or enable a later
  action.

Keep a candidate step merged only when the evidence is repeated observations of
the same primary action against the same primary target with the same security
intent.

A good step title should be narrow enough that it can be phrased as one verb
plus one object, for example:
- authenticate to an account;
- execute an attacker-controlled expression;
- inspect execution context;
- read a specific sensitive file;
- list a specific directory;
- collect a specific data target.

These examples describe action shapes only. They are not required answers for
this incident.

Before final JSON, silently run this checklist:
1. If a step mentions multiple targets, split it unless the logs show a single
   repeated action.
2. If a step mentions both execution and later file access, split it.
3. If a step cites evidence from clearly different time clusters or different
   paths, split it.
4. If an edge skips over an intermediate action that you output as a step,
   connect through the intermediate step instead.
5. If ATT&CK labeling is uncertain, keep the action step and use UNKNOWN or low
   confidence rather than merging the step away.
6. If an attack-relevant anchor in the evidence ledger is not cited by any step,
   add it as a step unless it is truly duplicate/background.
\end{lstlisting}
\caption{Effective anti-merge and granularity-calibration prompt block. The first six rules are appended by the runner before the frozen calibration addendum; the remainder specifies merge conditions, action-shape examples, and the final checklist.}
\label{fig:app-prompt-granularity-a}
\end{figure*}

\begin{figure*}[p]
\centering
\begin{lstlisting}[style=appprompt,firstline=1,lastline=57]
Visible-anchor decomposition profile: calibrated evidence-grounded behavior units

Use this profile for mixed benchmark inputs where evidence may be security
alerts, raw logs, application logs, network logs, process logs, or provenance
observations. The goal is to avoid assuming that every dataset contains alerts
while still preserving meaningful alert-level decomposition when alert cards are
present.

Calibrated granularity goal:
- Each attack step should represent one causally meaningful attacker behavior:
  one primary action, one primary target or object, and one security intent.
- Do not make one step per evidence card, one step per log line, or one step per
  timestamp unless the incident metadata says `step_unit` is `alert-node` and a
  distinct alert-like evidence card is itself the behavior unit. Evidence cards
  are observations; attack steps are behavior units.
- Also do not compress multiple different behaviors into a broad campaign phase.
  The right level is between raw telemetry and high-level narrative stages.

Fairness rule:
- Use only evidence cards present in the incident input.
- Do not assume any hidden list of important evidence IDs. The model must infer
  importance from visible evidence-card content only.
- Treat alert-style fields as useful visible clues, not as guaranteed labels.
- If no alert-style fields are present, do not invent alerts. Reconstruct steps
  from visible behavior in logs.

Step-unit calibration:
- If incident metadata contains `step_unit: "alert-node"`, treat each distinct
  alert-like anchor as a first-class candidate step. Do not collapse several
  different `alert_node_id`, alert names, compromised entities, or alert
  descriptions into one campaign phase. Merge only clear duplicates or repeated
  observations of the same alert behavior.
- If incident metadata contains `step_unit: "action-unit"`, reconstruct
  behavior-level action units: merge repeated telemetry for the same primary
  action and target, but split when action, target, actor context, or security
  intent changes.
- If `step_unit` is absent or unclear, use the evidence-grounded behavior-unit
  rule and keep distinct visible anchors separate when merging would hide an
  attacker action.

Visible anchor protocol:
1. First identify candidate anchors from the visible evidence.
2. If a card has alert-like fields such as `alert_name`, `alert_type`,
   `alert_node_id`, `description`, `compromised_entity`, or `severity`, treat it
   as a candidate alert anchor.
3. Otherwise, treat concrete observed behaviors as candidate anchors, including
   requests, commands, interpreter expressions, file reads/writes, process
   starts, authentication events, network connections, account usage, host
   changes, URLs, or sensitive resource accesses.
4. Create a separate attack step when anchors differ by primary action, primary
   target, account, host, URL, file path, process, or security intent.
5. Treat timestamp clusters as a split signal only when the later observation
   changes the primary action, target, actor context, or security intent. Do not
   split repeated observations merely because they happened at different times.
6. Merge anchors when they are duplicate, repeated, or supporting observations
   of the same primary action against the same primary target with the same
   security intent.
\end{lstlisting}
\caption{Visible-anchor decomposition prompt (part 1 of 2): cross-source step-unit calibration and split/merge protocol.}
\label{fig:app-prompt-visible-a}
\end{figure*}

\begin{figure*}[p]
\centering
\begin{lstlisting}[style=appprompt,firstline=58,lastline=102]
7. Do not create a new step for telemetry that only confirms, enriches, or
   reports another step, such as an alert firing for a command already captured,
   a process event confirming the same command, or repeated network logs for the
   same connection intent.
8. Prefer one direct anchor plus at most two directly supporting cards per step.
9. Do not cite a large bundle of loosely related evidence IDs for one step. If a
   step would need many evidence IDs because it covers different actions or
   targets, split it. If it needs several IDs only because they repeat the same
   behavior, keep one step and cite the strongest direct evidence.
10. If ATT&CK mapping is uncertain, keep the action step and set tactic,
   technique_id, or technique_name to `UNKNOWN` rather than deleting the step.

Coverage audit before final JSON:
- Silently make a table of observed anchors and assign each one to one of:
  `new_step`, `supports_existing_step`, `duplicate`, or `background`.
- Do not output the table.
- Every anchor marked `new_step` must appear as a separate attack_chain item.
- Every attack_chain item should cite the direct anchor evidence ID first, then
  only the strongest directly supporting IDs.
- If many observed attack-relevant evidence IDs are uncited, revise the chain
  before submitting rather than summarizing them away.

Split versus merge checklist:
- Split when the proposed step contains two different verbs, for example
  authenticate and execute, execute and read, discover and collect, or collect
  and exfiltrate.
- Split when the primary object changes in a security-relevant way, such as a
  different credential, sensitive file, account, host, URL, process, or data
  store.
- Merge when evidence IDs describe the same command/request/process/file access
  from different sensors or repeated log entries.
- Merge when several anchors differ only in wording, severity, rule name, or
  timestamp but point to the same primary behavior.
- Before final JSON, silently review the chain for over-fragmentation: if two
  adjacent steps have the same primary action, same primary target, and same
  security intent, consolidate them and keep the strongest evidence IDs.

Output contract:
- Return exactly one JSON object with root keys `incident_summary`,
  `attack_chain`, and `causal_edges`.
- Do not return `retrieval_report`, `analysis`, `attack_timeline`,
  `attack_stages`, anchor coverage tables, markdown commentary, or any
  diagnostic wrapper.
- Every attack_chain item must follow the required schema and cite only visible
  evidence IDs.
\end{lstlisting}
\caption{Visible-anchor decomposition prompt (part 2 of 2): citation discipline, coverage audit, and final output contract.}
\label{fig:app-prompt-visible-b}
\end{figure*}

\begin{figure*}[p]
\centering
\begin{lstlisting}[style=appprompt,firstline=1,lastline=57]
You are operating in AGENTIC attack-chain reconstruction mode.

You do not receive the full evidence package. Your working memory policy is
controlled by working_memory_mode.

Available actions:
- query_evidence: search for evidence supporting a missing or uncertain attack step.
- expand_entity: pivot from an observed host, account, file, process, IP, URL, or path.
- inspect_evidence: inspect specific evidence IDs and return their full currently available card fields.
- check_ordering: search for timestamps or causal/temporal links between candidate steps.
- submit: stop and submit working_chain as the final answer.
- give_up: stop when continued querying is unlikely to recover a defensible chain.

Return JSON only with this shape:
{
  "thought": "brief private-facing summary of what evidence is missing or why you are done",
  "action": "query_evidence|expand_entity|inspect_evidence|check_ordering|submit|give_up",
  "query_intent": "short natural-language retrieval query; empty only when submitting",
  "working_chain": {
    "incident_summary": "",
    "attack_chain": [],
    "causal_edges": []
  }
}

Agent constraints:
- Keep working_chain schema-compatible with the attack-chain JSON shape.
- Add or revise steps only when supported by observed evidence IDs.
- Prefer another query when key causal links, step evidence, or ordering are still uncertain.
- If retrieval/inspection stops revealing new evidence, synthesize the most defensible coarse-grained chain from observed telemetry; do not leave working_chain empty after evidence has been observed.
- Submit by turn 15; current turn is [TURN].
- Working memory mode: structured_attack_chain. Maintain working_chain incrementally every turn; after observing evidence, carry at least tentative low-confidence evidence-grounded steps when possible.
- Evidence memory compression: on; memory token budget is approximately 12000 tokens.
- Action mode: typed. Prefer query_evidence for missing support, expand_entity for entity pivots, inspect_evidence for a specific evidence ID/entity, and check_ordering for temporal uncertainty.
- Investigation memory is enabled. Use entity_memory and hypothesis_memory to plan targeted next queries.
- Support audit is enabled. Repair unsupported, unobserved, or weakly supported steps before submitting.

Final answer contract for working_chain:
[VERBATIM TASK CONTRACT AND VISIBLE-ANCHOR PROFILE FROM THE PRECEDING PANELS]

Incident metadata without full evidence:
[CASE-SPECIFIC VISIBLE INCIDENT STUB]

Current working_chain:
[CURRENT SCHEMA-COMPATIBLE WORKING CHAIN]

Current free_text_notes:
N/A

Investigation memory:
[RUN-LOCAL ENTITY AND HYPOTHESIS MEMORY]

Support audit:
[DETERMINISTIC SUPPORT-AUDIT SUMMARY]

Retrieved evidence memory:
[BOUNDED ECRAG OBSERVATIONS; VISIBLE EVIDENCE ONLY]
\end{lstlisting}
\caption{Per-turn controller prompt: typed actions, response schema, structured-memory constraints, and runtime-populated state. Uppercase bracketed fields are populated only from model-visible, run-local state.}
\label{fig:app-prompt-controller-a}
\end{figure*}

\FloatBarrier

\end{document}